\documentclass[journal,compsoc,10pt]{IEEEtran}

\usepackage[pdftex]{graphicx}
\usepackage[center]{caption}
\usepackage{url}
\usepackage{listings}
\newcommand{\code}[1]{{\fontsize{9}{9}\textsf{#1}}}

\newcommand{\figwidth}[0]{0.9\columnwidth}

\begin{document}

\author{Stijn Volckaert, Bjorn De Sutter, Koen De Bosschere, and Per Larsen
  \thanks{The authors thank the Agency for Innovation by Science and Technology
    in Flanders (IWT) and the Fund for Scientific Research - Flanders.}}

\title{Multi-Variant Execution of Parallel Programs}
\IEEEtitleabstractindextext{%
\begin{abstract}
Multi-Variant Execution Environments (MVEEs) are a promising technique to
protect software against memory corruption attacks. They transparently execute
multiple, diversified variants (often referred to as replicae) of the software
receiving the same inputs. By enforcing and monitoring the lock-step
execution of the replicae's system calls, and by deploying diversity techniques
that prevent an attacker from simultaneously compromising multiple replicae,
MVEEs can block attacks before they succeed.
Existing MVEEs cannot handle non-trivial multi-threaded programs because their
undeterministic behavior introduces benign system call inconsistencies in the
replicae, which trigger false positive detections and deadlocks in the
MVEEs. This paper for the first time extends the generality of MVEEs to protect
multi-threaded software by means of secure and efficient synchronization
replication agents. On the PARSEC 2.1 parallel benchmarks running with four
worker threads, our prototype MVEE incurs a run-time overhead of only 1.32x.
\end{abstract}

}

\maketitle
\IEEEdisplaynontitleabstractindextext

\section{Motivation}
Memory corruption vulnerabilities have been used to undermine the security of
computer systems for at least four decades. To this day, new attacks appear
shortly after new mitigations are introduced and long before they can be put
into practice. A clear trend is that transparent mitigations with low
overheads are able to enter practice. Address Space Layout
Randomization~\cite{PaXASLR}, Stack Cookies~\cite{cowan1998stackguard} and Data Execution
Prevention~\cite{PaXWX} are technologies that meet the high bar to
adoption. However, both the broadly deployed defenses and many emerging ones
have already been
bypassed~\cite{shacham2004effectiveness,richarte2002four,shacham2007geometry,goktas2014out,davi2014stitching,carlini2014rop,goktas2014size,schuster2014evaluating}.
Solutions that provide much stronger protection are less transparent
and involve more overhead. They include Software-based Fault
Isolation~\cite{wahbe1994efficient}, Control-Flow
Integrity~\cite{abadi2005control}, Code-Pointer
Integrity~\cite{kuznetsov2014code}, and Code
Diversification~\cite{larsen2014sok}.

The widespread availability of multi-core processors has made Multi-Variant
Execution Environments (MVEEs)~\cite{cox2006n,salamat2009orchestra} an
increasingly attractive solution. MVEEs monitor multiple instances, known as
replicae, of the same program running in parallel. When the MVEE detects
inconsistencies in the replicae's I/O behavior, it considers those as symptoms
of an ongoing attack and it terminates the replicae's execution to prevent harm.

The security guarantees provided by an MVEE build on three properties: (i)
isolation of the monitor from the replicae; (ii) monitored lock-step system call
execution; and (iii) diversification of the replicae. The isolation of the
monitor by means of hardware-enforced boundaries is achieved by implementing it
in kernel-space~\cite{cox2006n} or in a separate user-space
process~\cite{cavallaro2007comprehensive,salamat2009orchestra,maurer2012tachyon,hosek2013safe,volckaert2013ghumvee}.
All system calls invoked in the replicae are monitored, executed in lock-step,
and only allowed to execute when all replicae invoke the same system calls with
consistent inputs. This allows the monitor to detect when a single replica is
compromised and to halt its execution. However, this also implies that all
replicae only progress at the speed of the slowest one. Furthermore, it implies
that the monitor cannot look ahead at future system calls when deciding if a
specific system call invocation should be allowed. This places strict
constraints on the system calls in the replicae. Foremost, they need to occur in
the same order in all replicae.

Those replicae are diversified instances of the same program. Several techniques
are suitable to generate the diversified replicae, including include Address
Space Partitioning~\cite{cavallaro2007comprehensive}, Reverse Stack
Growth~\cite{salamat2008reverse}, System Call Number
Randomization~\cite{chew2002mitigating} and Disjoint Code
Layout~\cite{volckaert2015cloning}. All of these techniques aim to introduce
enough diversity to cause detectable divergences in the system call behavior of
replicae under attack, while maintaining consistent system call behavior if the
inputs are benign. This task is complicated by the fact that the system call
behavior does not solely depend on explicit, user-provided program inputs, but
might also depend on implicit inputs.

We prevent benign system call consistency violations that result from implicit
inputs, such as randomized virtual addresses that affect the system call
behavior through address-dependent computations, by using \emph{implicit-input
  replication agents}~\cite{volckaert2013ghumvee}. These agents log the implicit
inputs in one replica, the master, and replicate those inputs in the other
replicae, the slaves, thus restoring system call consistency. Together with the
system calls, the points at which these agents intervene form the so-called
\emph{rendez-vous points} (RVPs) of the replicae.

In this work, we focus on a big weakness of existing secure MVEEs with respect
to system call consistency: secure replication of non-deterministic
multi-threaded applications. In real-life multi-threaded programs, even the
security-sensitive system calls that should be monitored most strictly often
differ between replicae as a result of their non-deterministic
scheduling. Because of the lock-step system call execution and monitoring
requirement, security-oriented MVEEs cannot tolerate those divergences.

Two broad classes of techniques could potentially alleviate this problem. First,
a Deterministic Multi-Threading (DMT) system can be embedded in the program to
enforce a fixed thread schedule in all
replicae~\cite{basile2002preemptive,reiser2006consistent,berger2009grace,liu2011dthreads,merrifield2013conversion,cui2013parrot,olszewski2009kendo,lu2014efficient,devietti2009dmp,bergan2010coredet,zhou2012exploiting}. In
the context of an MVEE, however, all existing DMT systems fall short. To
establish a deterministic schedule, they all rely, in one way or another, on a
token. Some systems only allow threads to pass this token when they invoke a
synchronization operation. This approach is incompatible with threads that
deliberately wait in an infinite loop for an event such as the delivery of a
signal to trigger because such threads may never invoke a synchronization
operation.  Other systems allow threads to pass their token when they have
executed a certain number of instructions. Such systems cannot tolerate
variations in the program execution and are therefore incompatible with most
code diversification techniques as well as the implicit-input replication
agents.

Alternatively, we can accept non-determinism and require only that all replicae
execute in the same non-deterministic order. Online Record/Replay (R+R) systems
can provide this guarantee by logging the execution in one replica and replaying
it in the other
replicae~\cite{basile2006active,lee2010respec,basu2011karma}. R+R systems are
less sensitive to variations in the program execution, which we typically see
with diversified replicae. But in order to use them in a security-oriented MVEE,
they need to be adapted to become address-agnostic and to support programs that
use ad hoc (i.e. non-standardized) synchonization primitives or lock-free algorithms. Furthermore, for
embedding an R+R system in a security-oriented MVEE we need to ensure that any new
functionality introduced in the replicae must be neutral with respect to the
RVPs, and we need to secure the communication channel that is used to convey the
information about the recorded execution from the master replica to the slave
replicae.

Our paper makes four contributions. First, we present four R+R-based
synchronization replication agents that record synchronization operations in a
master replica and replay them in the slaves, thus ensuring system call
consistency. The agents are address-agnostic and system-call-neutral, and hence
compatible with existing secure MVEEs and implicit-input replication agents. The
most efficient agent communicates over a channel that is secured against
malicious communication by attackers.

Second, we present a practical strategy to extend our R+R-based systems to support
  ad hoc and lock-free synchronization, which we typically see in many low-level
  libraries. 

Third, we report how we integrated our replication agents into GNU's \emph{glibc}
  and how we applied the aforementioned strategy to four commonly used
  system libraries: GNU's \emph{libpthreads}, \emph{libstdc++} and
  \emph{libgomp}. This integration enables support for data
  race free C and C++ programs that use the \emph{pthread} and/or \emph{OpenMP}
  programming models.

Finally, we extensively evaluate the run-time performance of our replication
agents, the implementation effort that went into their integration into the
aforementioned libraries, and the security of the proposed features.

\section{Replication of Multi-Threaded Programs}
\label{replication}

The techniques presented in this paper build on GHUMVEE, an existing
security-oriented MVEE~\cite{volckaert2013ghumvee,volckaert2015cloning}. To
monitor the replicae, GHUMVEE uses the \code{ptrace} and \code{process\_vm\_*}
Linux APIs. As the use of these APIs involves context switching, they introduce
significant latencies in the interaction between the monitor and the
replicae. This makes them unacceptable for replicating synchronization events,
which occur frequently in many programs and which are often handled entirely in
user space in the original programs to optimize performance. For example, we
observed \code{gcalctool}, a simple calculator from the GNOME desktop
environment performing 1.8M futex operations during its 400 ms initialization,
almost all of which were uncontended and hence handled in user space.
Interposing all those operations with system calls and ptrace made the
initialization time grow to over 370 seconds, a slowdown with a factor 925!

To avoid such an unacceptable overhead, our alternative solution consists of a
synchronization replication agent that replicates all synchronization events
entirely in user space.

\subsection{Synchronization Replication Agent}
\label{repagent}

\begin{figure}[t!]
\centering
\includegraphics[width=\figwidth]{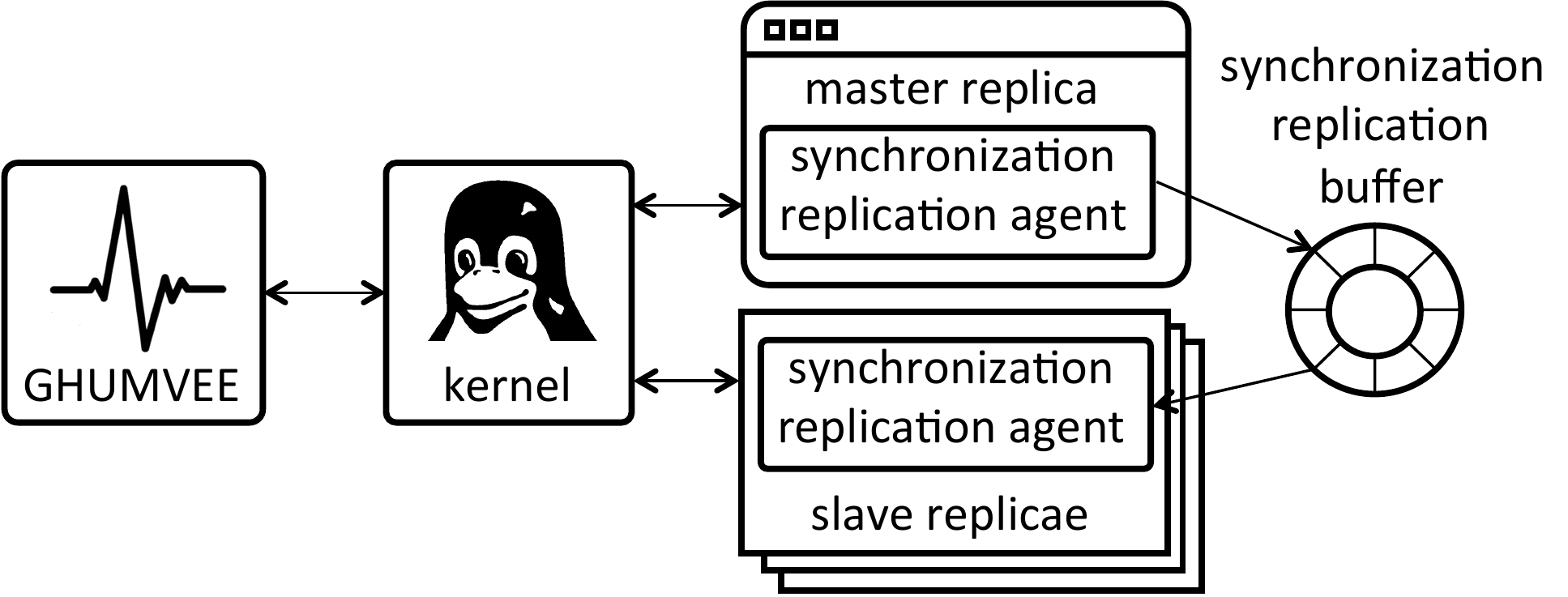}
\caption{System overview}
\vskip -0.3cm
\label{fig:system_overview}
\end{figure}

We enforce an equivalent execution in all replicae by injecting a
synchronization replication agent into their address space, as shown in
Figure~\ref{fig:system_overview}. At run time this agent forces the master
replica to capture the order of all inter-thread synchronization operations,
hereafter referred to as \emph{sync ops}. The agent logs the captured order in a
circular, shared buffer that is visible to all replicae. This buffer is mapped
with read/write permission in the master replica and with read-only permission,
and at different addresses, in the slave replicae. In the slave replicae, the
agent uses the captured order to enforce an equivalent replay of sync ops.

To capture the sync op execution order, we wrap them in a small critical
section at the source code level. Within the critical section
we first log information about the sync op in the first available slot of the
buffer and then perform the original op. Depending on the
agent we use, the information about the sync op consists of the
thread ID, the memory word that was affected by the op, and the op's type. 

The replication agent must be available to the entire program, including any
loaded shared libraries. So we chose to implement the agent in
\code{glibc}, at the lowest possible level in the user-mode portion of the
software stack where it is exposed to the program itself and to all shared
libraries.

\begin{figure}[t!]
  \centering
  \begin{tabular}{c}
    \includegraphics[width=\figwidth]{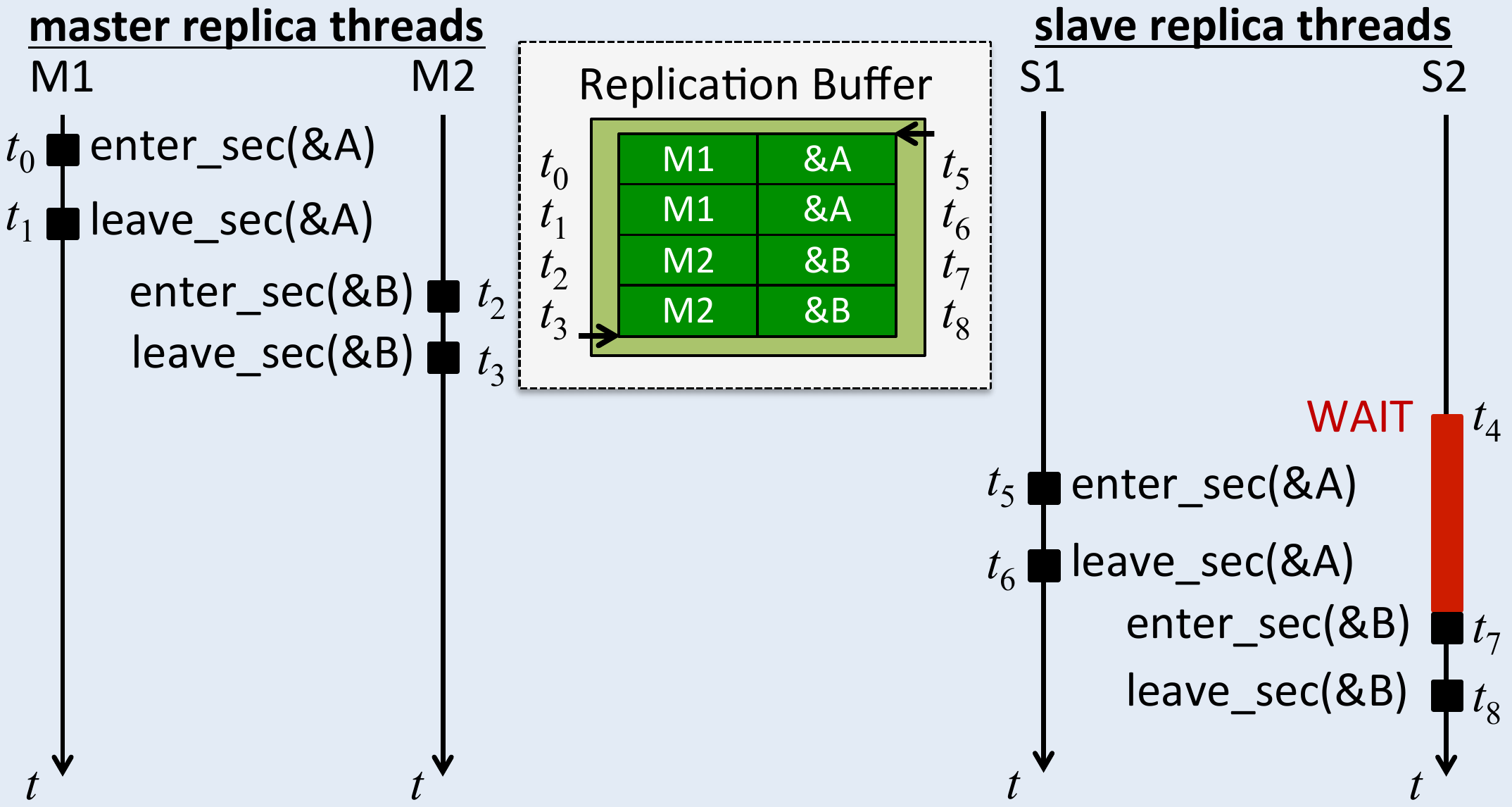}\\[-3pt]
    \small (a) Total-order replication\\[6pt]
    \includegraphics[width=\figwidth]{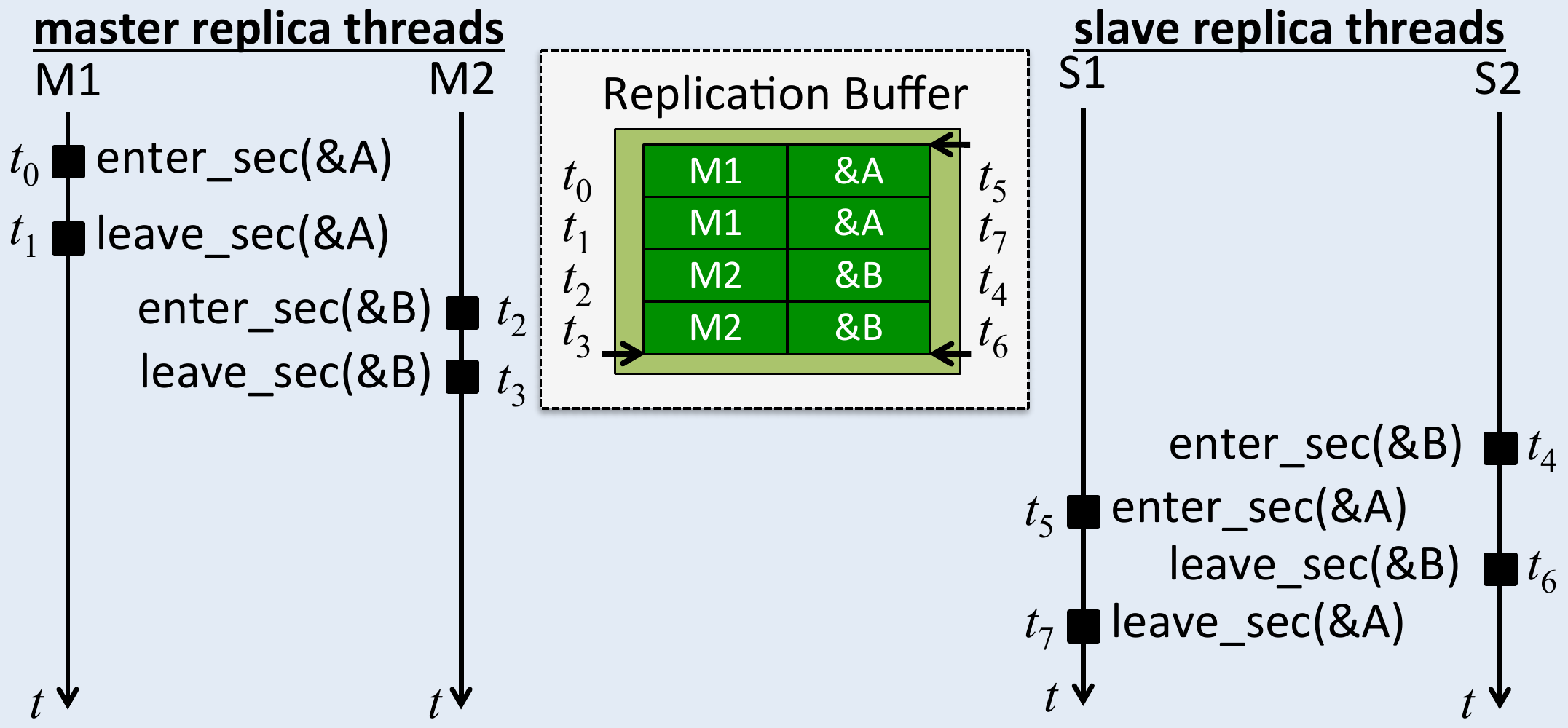}\\[-3pt]
    \small (b) Partial-order replication\\[6pt]
    \includegraphics[width=\figwidth]{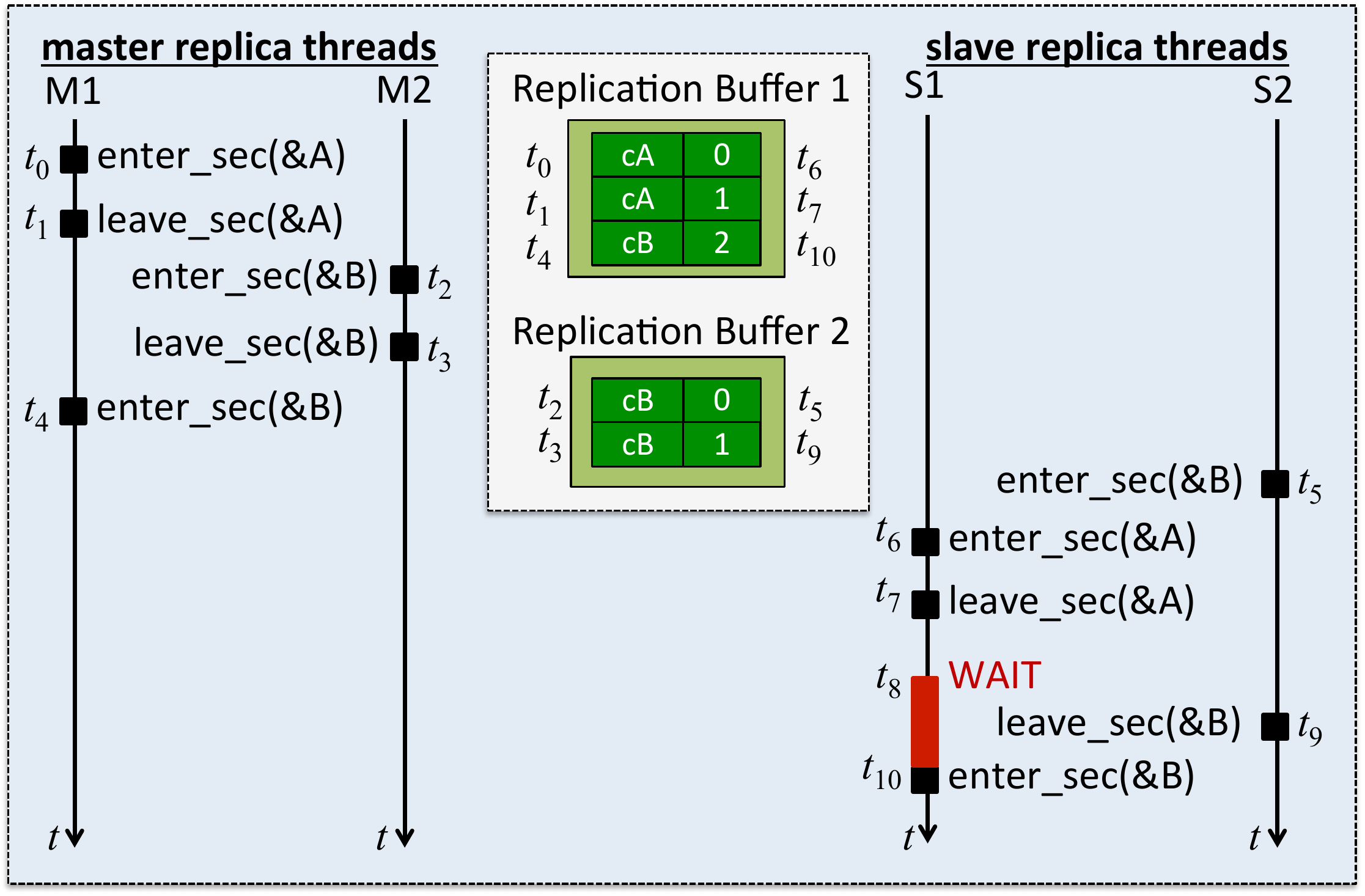}\\[-3pt]
    \small (c) Wall-of-clocks replication\\[-7pt]
  \end{tabular}
\caption{Replay sequences with three replication strategies}
\label{figorder}
\end{figure}

The same agent is used in the master and slave replicae: Identical instances of
\code{glibc} are loaded into the master and slave replicae when they are
launched, though they might be loaded different addresses in each replica. When
an instance is later invoked at run time, it needs to know whether it is invoked
in a master or slave replicae, to either capture or replay the sync op order. We
therefore dynamically initialize the agent in each replica. Soon after a replica
is launched by the MVEE monitor, its agent invokes a system call that is
intercepted by the monitor. Through this system call, the agent passes the
location of its status flags to the monitor. At that point, and at each later
intervention from the monitor in the replicae, the monitor can configure the
agent instance as a master or slave agent. Through the status flags, the monitor
can also disable the agent when the replicae are executing a single thread, and
enable the agent when the replicae (are about to) start executing multiple
threads. From that initial configuration onwards, the agent in each replica
communicates only with the agents in the other replicae. With the exception of
being enabled/disabled, the agents do not communicate with the monitor. This
avoids the extensive context switches that would result from using the
\code{ptrace} or \code{process\_vm\_*} APIs for replication.

While the high-level principles of our replication agents are reminiscent of
online R+R techniques, we cannot trivially adopt them. Within an MVEE, it is
critical that any new functionality injected into the original code is
\emph{neutral with respect to RVPs}. Specifically, we have to ensure that if the
new functionality introduces a new RVP, this RVP is introduced in all replicae
at the exact same point in the program execution and with equivalent
arguments. Furthermore, because a secure MVEE needs to enforce lock-step
execution on the replicae, we need to replicate information actively and with
minimal delay to avoid that slave replicae delay the master replica too much. We
therefore implement replication agents that make the recorded information
\emph{visible immediately}, rather than broadcasting it periodically.

These two design decisions have far-reaching consequences. First, the
RVP-neutrality constraint prevents us from using dynamic memory allocators,
because those use sync ops to coordinate multi-threaded access to the memory,
and introduce system call RVPs to allocate additional memory pages and to change
protection flags.

Second, since we want any information to be visible immediately in the slave
replicae, the agent cannot perform any post-processing on the recorded
information. This prevents us from compressing the recorded information to
reduce our agents' memory bandwidth requirements.

Third, our agent must support \emph{diversified replicae}. It can
therefore not assume that the master and slave replicae are fully identical. For
example, the same mutex might be found at different addresses in the different
replicae. Consequently, the recording side of the replication agent must record
its information in a manner that is \emph{address-agnostic}.

\subsection{Replication Strategies}
\label{strategies}

To replay the sync ops in the slave replicae, several approaches are available
that trade CPU cycles off against memory pressure. We have implemented three
replication strategies that meet the aforementioned constraints.

\subsubsection{Total-order replication agent}

Our total-order (TO) replication agent replays all sync ops in the exact same order
in which they happened in the master replica. Figure~\ref{figorder}(a) shows two
threads that execute under GHUMVEE's control.  In the master replica, thread
\code{M1} first enters and leaves a critical section protected by lock \code{A}
at times $t_0$ and $t_1$ resp.\ At those times, the wrappers of the
corresponding sync ops log the activities of thread M1 in the replication
buffer. Next, thread \code{M2} in the master replica enters and leaves a
critical section protected by lock \code{B} at times $t_2$ and $t_3$ resp. These
events are also logged in order in the buffer. Right after $t_3$, the buffer
holds the contents indicated in the figure. Time stamps to the left and right of
the buffer mark the time the buffer elements are produced and consumed resp. The
arrows on the left and right denote the position of the producer and the
consumer pointers resp.\ right after $t_3$.

In the slave replica thread \code{S2}, corresponding to \code{M2} in the master
replica, reaches the critical section protected by lock \code{B} first, at time
$t_4$. At that time, the first element in the buffer indicates that
synchronization events in the master replica occurred in thread \code{M1} first,
so thread \code{S2} is stalled in the wrapper of the sync op in
\code{enter\_sec}. Only after the first two elements in the buffer are consumed
in thread \code{S1} at times $t_5$ and $t_6$, can thread \code{S2} continue
executing. Thus, even though the two critical sections protected by locks
\code{A} and \code{B} are unrelated, thread \code{S2} is forced to stall until
thread \code{S1} has replayed the operations performed by thread \code{M1}.

This agent is trivial to implement, but not very efficient: The lack of
lookahead by consumers introduces unnecessary stalls as indicated by the red bar
in Figure~\ref{figorder}(a).

\subsubsection{Partial-order replication agent}

Our partial-order (PO) replication agent is more efficient. It only enforces a total
order on dependent sync ops. This agent may replay independent sync ops in any
order, as long as it preserves sequential consistency within the thread. The
PO agent is more complex and introduces more memory pressure because
the agents in the slave threads have to scan a window in the buffer to look
ahead. However, it typically introduces much less stalling and generally
outperforms the TO agent. In Figure~\ref{figorder}(b), we see the exact
same order of events as in Figure~\ref{figorder}(a) until $t_4$. This time,
however, thread \code{S2} may enter the critical section without delay at $t_4$
because the \code{enter\_sec} operation does not depend on either of the
operations that preceded it in the recorded total order.

Conceptually, there are significant similarities between this agent and online
R+R techniques such as LSA~\cite{basile2006active} and offline R+R techniques
such as RecPlay~\cite{ronsse1999recplay}. However, our agent captures events at
a finer granularity of \emph{sync ops} instead of \emph{pthread}-based
synchronization operations. Furthermore, our agent is fully RVP-neutral. These
are relatively minor differences, however, as techniques like LSA can be adapted
to capture at a lower granularity and to use only statically allocated memory. A
more fundamental difference is that our agent also supports diversified
replicae. With queue projection, LSA discards the per-thread order of
synchronization operations and only maintains the per-variable order. With
diversified replicae, the same logical variable might be stored at different
addresses in different replicae. Our agent therefore relies on the per-thread
order to determine which logical variable is affected by each synchronization
operation.

Altough the PO agent eliminates unneccessary stalling, it still
suffers from poor scalability. The master replica must
safely coordinate access to the circular buffer by determining the next free
position in which it can log an operation. If many threads simultaneously log
synchronization events, this inevitably leads to read-write sharing on the
variable that stores the next free position.  A similar problem exists on the
slave replicae's side because they must keep track of which data has been
consumed. With multiple slave replicae, this also leads to high sharing and,
consequently, high cache pressure and cache coherency traffic.  

\subsubsection{Wall-of-clocks replication agent}

The above observation led us to the design a third agent. This wall-of-clocks (WoC)
agent assigns each distinct memory location that is ever involved in a sync op
to a logical clock. These clocks capture ``happens-before'' relationships
between related sync ops~\cite{lamport1978time}. Similar to, e.g., plausible
clocks, but without using clock vectors, our clocks only capture the necessary
relationships~\cite{torresrojas1999plausible}.

In Figure~\ref{figorder}(c), lock \code{A} stored at address \code{\&A} is
assigned to clock \code{cA}. Lock \code{B} is similarly assigned to clock
\code{cB}.

On the master side, the agent logs the identifier of the logical
clock associated with each sync op, as well as that clock's time. After
logging each sync op, the agent increments the logical clock time of the
associated clock. 

In this agent, the logging is no longer done in a single circular
buffer. Instead there is one circular buffer per master thread, such that each
buffer has only one producer. In Figure~\ref{figorder}(c), master thread
\code{M1} only communicates with slave thread \code{S1} through buffer 1,
whereas thread \code{M2} only communicates with thread \code{S2} through buffer
2. This design avoids the contention for access to the shared buffers.

Neither the master nor the slave replicae need to propagate
their current buffer positions to other threads. Furthermore, the master's
logical clocks do not need to be visible to the slaves. The information
contained within the circular buffers suffices for the slave replicae to replay
the same clock increments on their own local copies of each clock.

In Figure~\ref{figorder}(c), thread \code{M1} first enters a critical section
protected by lock \code{A} at time $t_0$. The agent observes that the current
time on logical clock \code{cA} is 0. It records the clock and its time in
buffer 1 and increments the clock's time to 1. At time $t_1$, the agent logs the
exit from the critical section in buffer 1. This time around, the logical clock time is 1.

A similar situation then unfolds in thread \code{M2} at time $t_2$. This time
though, the critical section is protected by lock \code{B}, of which the
associated memory location is assigned to clock \code{cB}, whose initial time
also is 0. This information is logged in circular buffer 2, along with
information regarding the exit of the critical section in thread \code{M2} at
time $t_3$. At that point, clock \code{cB} is incremented to 2.

In thread \code{M1}, a third critical section is entered at time $t_4$,
which is again protected by lock \code{B}. This event involving logical clock
\code{cB} is logged in buffer 1 with clock time 2.

On the slave replica's side, the threads are scheduled differently in our
example. There, thread \code{S2} reaches a sync op first, at time $t_5$. The
agent observes in buffer 2 that it must wait until clock \code{cB} reaches time
0.  Since this is the initial time on the slave's copy of that clock, the
operation can be executed right away and thread \code{S2} will increment the
time on its copy of \code{cB} to 1. If we suppose that thread \code{S2} is then
pre-empted and thread \code{S1} gets scheduled, \code{S1} will enter and
leave the critical section protected by lock \code{A} at times $t_6$ and $t_7$,
consuming the first two entries in buffer 1, thereby incrementing the slave copy
of clock \code{cA} to 2.

The third operation in thread \code{S1} at time $t_8$ is the most
interesting. In the first replication buffer, the slave agent observes that the
sync op to enter a critical section has to wait until its associated logical
clock \code{cB} has reached time 2. However, in the slave, that clock's time was
last incremented at time $t_5$, i.e., to the value of 1. Thread \code{S1} must
therefore wait until some other slave thread has incremented the time on
\code{cB}. This will happen at time $t_9$ in thread \code{S2}. Shortly
thereafter, the agent code executing in thread \code{S1} will observe that
\code{cB} has reached the necessary value, and at $t_{10}$ \code{S1} will enter
its second critical section.

With this WoC, the replication agent only inserts accesses to shared
data, and hence coherence traffic, for two reasons. First, it introduces
accesses to replication buffers shared between corresponding threads in the
master and slave replicae. This is a fundamentally unavoidable form of overhead
required to replicate the synchronization behavior from the master to the slave
replicae. 

Secondly, the agent inserts accesses to shared clocks whenever multiple threads
in the original program were already contending for locks at shared memory
locations. While these extra shared accesses in the replication agents still
introduce some overhead, we do expect the overhead to scale with the
pre-existing resource contention in the original application. In other words, if
the original application uses contended global locks that decrease the available
parallelism, the replication agent will hurt it further. However, if the
original application involves a lot of synchronization, but that synchronization
is performed using uncontended local locks, the WoC replication agent
will not introduce contended traffic within the master or slave replicae either.

As we will see in Section~\ref{evaluation}, the WoC agent
consistently outperforms the other agents on almost every benchmark. Most
importantly, as is the case with plausible clocks in general, the replication
will always be correct~\cite{torresrojas1999plausible}.

One important remark remains to be made, however. While the WoC agent
is certainly the more elegant and more efficient of the three proposed designs,
it is not fully optimal. Due to the RVP-neutrality constraint, we cannot
dynamically assign each memory location to its own private clock. Instead, we
have to pre-allocate a fixed number of clocks statically and we have to assign
lock memory locations to one of those clocks based on a hash of their memory
address. Because we want to use a cheap hash function, hash collusions are quite
likely. Any such collusion results in an $m$-to-1 mapping between multiple locks
and each clock. In other words, the WoC agent is bound to assign some
non-conflicting memory locations to the same logical clock. When this happens,
this introduces unnecessary serialization and hence potentially also
unnecessary stalls in the slave replicae.

Our WoC agent is similar to Respec~\cite{lee2010respec}, although it
does not share any part of its implementation. It differs from other clock-based
techniques, however, in that it does not use thread clocks. Instead, our agent
relies solely on the logical clock it assigns to each memory location. In the
ideal case, our agent therefore only needs to read and update the value of one
clock to replay a synchronization operation. Techniques that rely on Lamport
clocks (e.g. ROLT~\cite{levrouw1994new}) by contrast need to read and update the
values of two clocks: the local thread clock and the synchronization variable's
associated clock. Techniques that rely on vector clocks
(e.g. RecPlay~\cite{ronsse1999recplay}) need to read the value of at least $n+1$
clocks (with $n$ the number of threads in the program): the local thread clock,
the synchronization variable's clock, and the thread clocks of all other
threads.  The reason why our agent does not need local thread clocks is that it
records into a per-thread buffer, rather than a globally shared
buffer. Therefore, a thread clock would never have to be synchronized with other
thread clocks, which eliminates the need for such a clock
altogether. Furthermore, the fact that our agent assigns each memory location to
a statically allocated clock implies that the agent can be applied transparently
and that it respects RVP-neutrality.

\subsection{Secured wall-of-clocks agent}
\label{swoc}

\begin{figure}[t!]
\centering
\includegraphics[width=7.5cm]{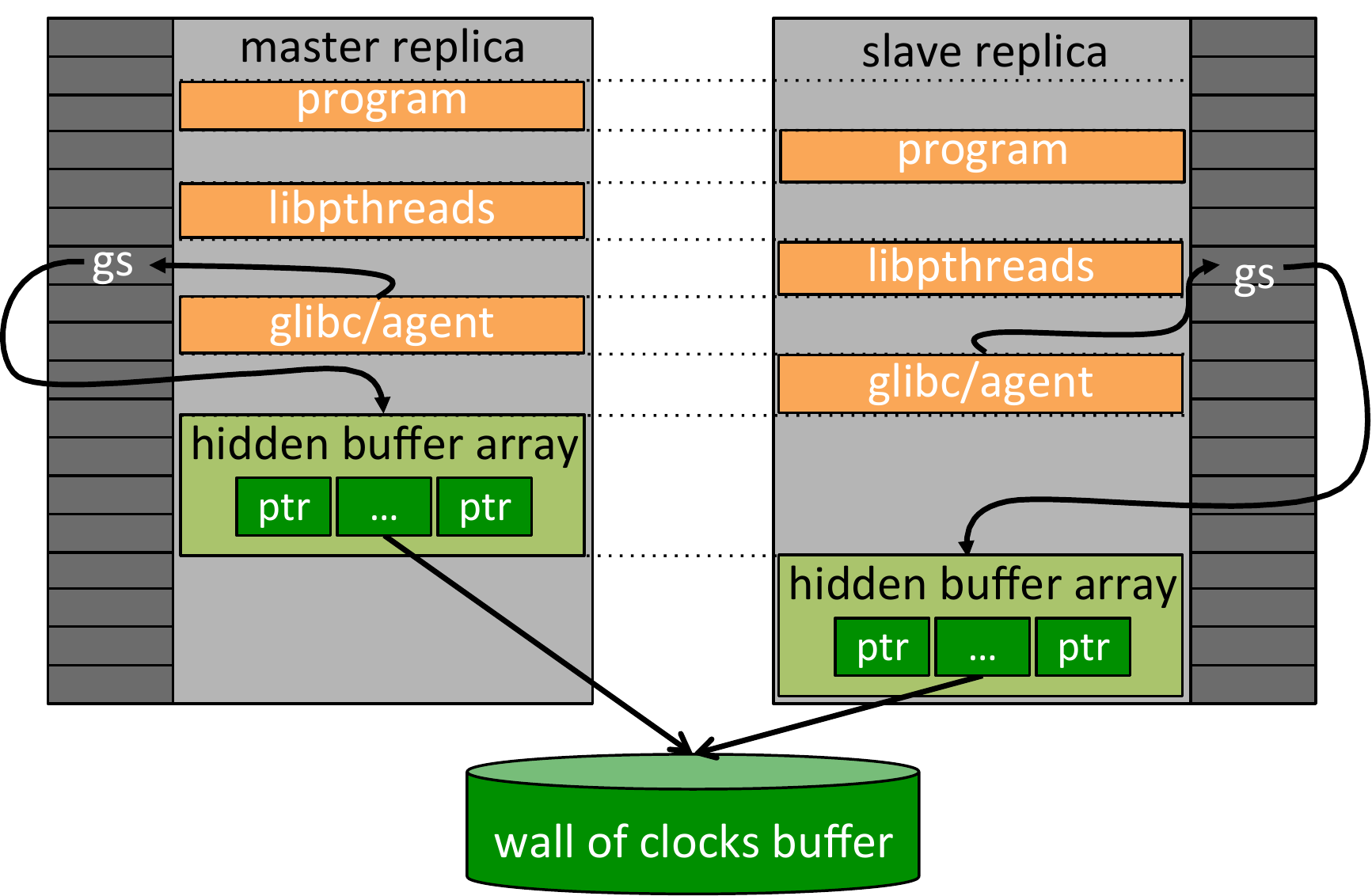}
\caption{Hidden buffer array access to the replication buffer.}
\label{fig:HBA}
\end{figure}

The agents implementing the three replication strategies as discussed in the
last sections are not very secure: They forward information through a circular
buffer that is shared among all replicae. This buffer easy to locate since all
three of these agents store a pointer to it in a thread-local variable. Despite
of the code reuse countermeasures we have in place~\cite{volckaert2015cloning},
attackers could exploit the fact that an easily locatable communication channel
between the replicae exists to set up an attack that can compromise multiple
replicae. 

As the WoC agent outperforms the other two agents on average, we
build on that agent to present an alternative, secured design that relies on the
\emph{hidden buffer array} (HBA) shown in Figure~\ref{fig:HBA}. This page-sized
array stores pointers to hidden buffers.  Upon startup, a replica can request
that this HBA be allocated by GHUMVEE and subsequently map it into its own
address space using the System V IPC API~\cite{sysv}. GHUMVEE intercepts and
manipulates this mapping call such that the pointer to the HBA is not returned
to the program. At the same time though, GHUMVEE overrides the base address of
the replica's \code{gs} segment so that it points to the HBA.

The reason to override this address is that the \code{x86\_64} architecture
supports addressing of 48-bit (or bigger) pointers and has therefore disabled
most of the original \code{x86} segmentation functionality. The \code{gs} and
\code{fs} segment registers may still be used as additional base registers,
however, and by consequence all \code{gs} or \code{fs}-relative memory accesses
are still valid. It is extremely uncommon to still find such accesses in
\code{x86\_64} software, however. Furthermore, \code{x86} processors do not
allow user-space instructions to read the segment registers. The
\code{gs} and \code{fs} segments can therefore be used to store pointers that
are hidden from the user-space software.

At a fixed offset within the HBA we store a pointer to the agent's circular
buffer. The end result is that the replica must read the pointer to the circular
buffer indirectly, through a \code{gs}-relative memory access. In assembler, we
manually crafted a version of our WoC agent that accesses this
pointer in such a way. By storing the pointer to the buffer and any pointers
derived from it in a fixed caller-saved general-purpose register, we guarantee
that the pointer never leaks to memory and that no function outside the
replication agent can observe the pointer value. We further
guarantee that (i) the pointer to the buffer is never moved to a different
register, (ii) the register is never pushed onto the stack, (iii) the register
is cleared before the function returns and (iv) the replication agent does not
call any functions while the pointer value is visible.

Since neither \code{gcc}, nor \code{LLVM} offers syntactic sugar to allow for
such properties, we have chosen to implement both of the replication agent's
functions that access the shared buffer in assembly code. The current
implementation, which we evaluate in Section~\ref{evaluation}, totals
approximately 150 LoC.

\subsection{Embedding the replication agent}
\label{rdready}

The key challenge in embedding the replication agent into a program is to
identify all sync ops. Because we want to wrap these sync ops in the source code
itself, we also identify the sync ops at the source code level.

Existing R+R systems, as well as DMT systems that impose weak determinism, only
order invocations of \code{pthread}-based synchronization functions. That is
insufficient for a secure MVEE because
\code{glibc} and several other low-level libraries implement their
own sync ops. A failure to order the sync ops in \code{glibc} tends not
to affect the user-observable I/O determinism of the program, but it does
impact the general system call behavior and hence violates the system call
consistency needed in a secure MVEE.

An alternative strategy would be to order all sync ops by wrapping all loads and
stores in the program. This would yield system call consistency even in the
presence of data races. Ordering individual loads and stores leads to
prohibitively high overhead however, as was demonstrated in the context of
strong determinism~\cite{devietti2009dmp}. Moreover, given the range of
diversity we need to support in the replicae to mitigate a sufficiently wide
range of attacks, there is no guaranteed one-to-one mapping between the loads
and stores in the different replicae.
  
\begin{figure}[t!]
\centering 
\includegraphics[width=\columnwidth]{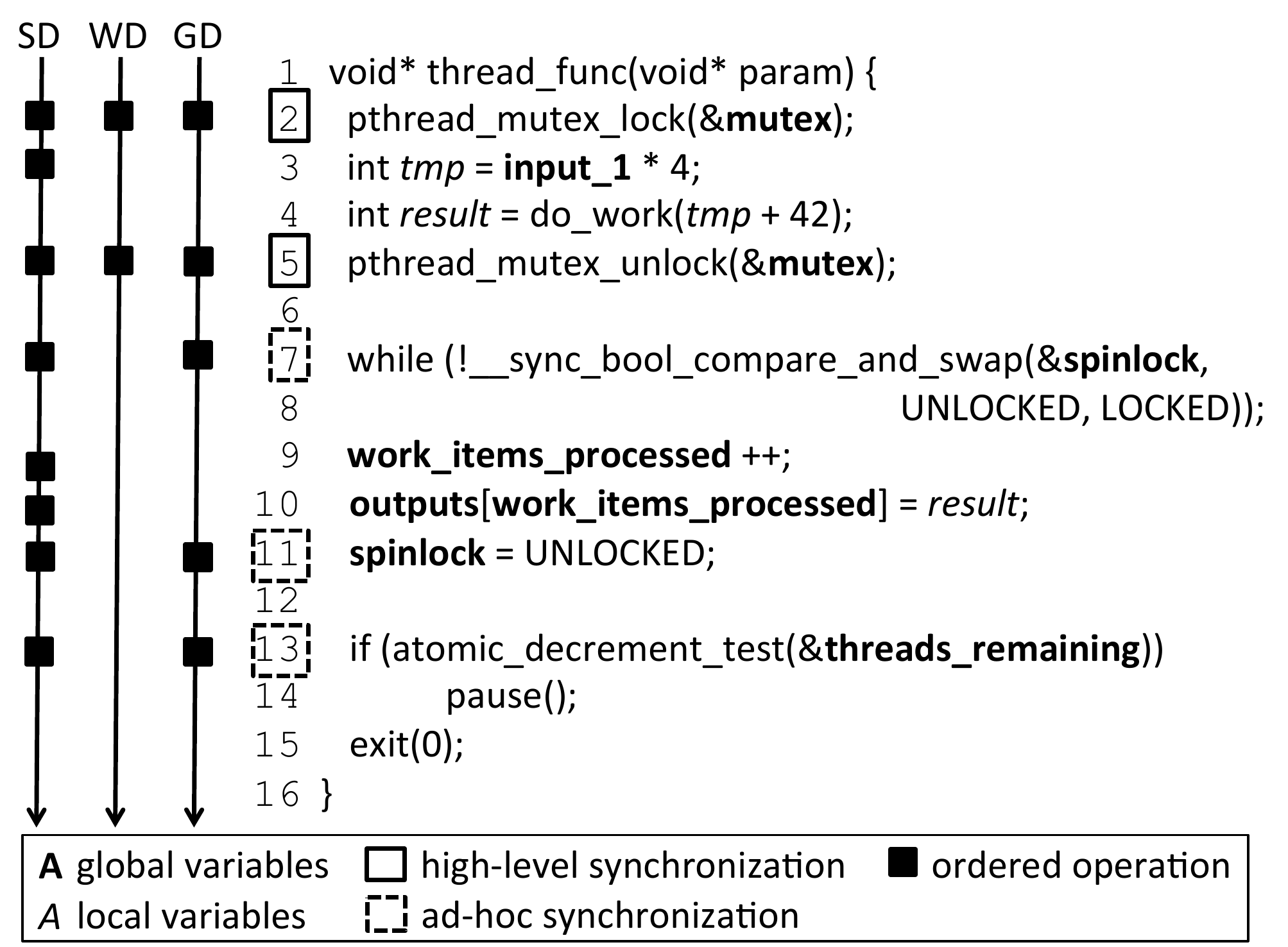}
\caption{Comparison between three classes of determinism.}
\label{comparison_determinism}
\end{figure}

The strategy we propose is most similar to weak determinism systems, but we
capture sync ops at a lower level as shown in
Figure~\ref{comparison_determinism}. In existing strong determinism systems, all
11 memory operations need to be ordered. (Lines 3, 9, and 10 each involve two
memory operations.) With existing systems of weak determinism, only the two
(standard synchronization) operations on the mutex on lines 2 and 5 are
ordered. In our solution, we wrap both the standard operations on the mutexes as
well as the three ad hoc synchronization events on lines 7, 11 and 13. The
latter one translates into a \code{LOCK SUB} instruction on the x86
architecture, and atomically sets the zero flag (ZF).

To identify the source lines to be wrapped, we first identify the sync
ops in the binary code, and then translate those to source line numbers by means
of debug information. 

The relevant sync ops in the binary code come in three categories. First, any
instructions that the programmer explicitly marks as atomic are sync ops. On the
x86 architecture, this includes all instructions with an explicit \code{LOCK}
prefix, as well as \code{XCHG} instructions with an implicit \code{LOCK}
prefix~\cite{IntelManual}. \code{LOCK CMPXCHG} is an example. Second, any store
operation (e.g., for a C assignment to a dereferenced pointer or volatile
variable) that directly succeeds an explicit memory barrier is a sync op. Such
stores are typically used in synchronization schemes like read-copy-update
(RCU). Third, any instruction that references a memory-address (such as some
memory-allocated variable) that is referenced by another sync op also becomes a
sync op. We refer to these operations as \emph{unprotected loads and stores}, a
terminology sometimes used to denote benign data races.
  
The guarantees we provide by running a replication agent that enforces an
equivalent order of sync ops in all replicae are at least as strong as the
guarantees that weak determinism provides. To see why, recall that weak
determinism enforces a deterministic order of entries into critical
sections. Thus, in a correct program, weak determinism will grant mutually
exclusive access to related blocks of shared memory in a deterministic
order. Though this was historically not the case~\cite{dijkstra1965solution},
all modern user-mode programs that run on SMP-systems now implement mutual
exclusion using an atomic test-and-set
operation~\cite{lamport1987fast,mellor1991algorithms}. On x86 systems, several
instructions provide test-and-set semantics, but they are atomic only if a
\code{LOCK} prefix is used. Consequently, the first category contains all
instructions that may implement the entrance into a critical section as a sync
op. Operations that implement the exit from a critical section can either be
implemented using atomic test-and-set or exchange operations (second category)
or with atomic store operations (first or third category).

On the GNU/Linux platform, all high-level synchronization primitives in the
\code{pthreads}, \code{OpenMP} and \code{C++} standard libraries are based
on the mutual exclusion principle. In the \code{pthreads} library, e.g.,
functions such as \code{pthread\_mutex\_lock} and \code{pthread\_cond\_wait}
implement mutual exclusion using \code{LOCK CMPXCHG} instructions. In the
implementation of other high-level synchronization primitives, a variety of
atomic test-and-set operations are used. All of them are prepended with
\code{LOCK} prefixes, however, and all of them are therefore classified as
sync ops.

The identification and wrapping of sync ops is currently a partially manual
process. First, we disassemble the binary/library to identify explicit memory
barrier instructions or instructions with explicit or implicit \code{LOCK}
prefixes. If no such instructions are present, no further steps are needed. We
use debugging symbols to map all of the identified instructions to their
originating source line. For memory barriers, we wrap the store that directly
succeeds the barrier in calls to our replication API. We identify loads of the
same variable and wrap them too. To some extent, we thus fix benign and
deliberate data races, such as when developers use such a barrier to set a
flag without synchronization.
  
For source lines that compile into instructions with \code{LOCK} prefixes, check
whether or not compiler intrinsics are used to express the atomic operations. If
not, we insert calls to our replication API before and after the
operation. Otherwise, we include an automatically generated header in the source
file. This header overrides all known intrinsics that implement atomic
operations and inserts the appropriate replication API calls automatically. It
is generated by a simple script that downloads the list of all atomic compiler
intrinisics from the GNU GCC website, parses the list and generates the
necessary definitions. In addition, we identify other loads and stores of the
variables involved in the atomic operations and insert API calls manually.
  
While this process might seem cumbersome, it is important to note that unless a
program or higher-level library implements its own ad hoc synchronization or
lock-free algorithms, no memory barriers or instructions with LOCK prefixes will
be found. So most programs are supported transparently. For our tests, only four
libraries needed modifications. We report extensively on the size of these
modifications in Section~\ref{eval_ready}.

In many cases, and in particular in portable code, the manual effort to invest
in these modifications is very limited. In portable code, compiler intrisics are
used to implement atomic operations, as well as all other accesses to the
variables involved in those atomic operations. This is necessary to ensure
portability to architectures that do not guarantee the atomicity of aligned
loads and stores. To wrap all the necessary compiler intrinsics, it suffices to
include our automatically generated header in all source files.

Furthermore, even for programs and libraries that do need more modifications,
the patching process can be streamlined to a great extent. C++11 compliant
compilers provide a template for atomic operations~\cite{Catomic}. With this
template, a programmer can mark variables that need to be updated atomically by
modifying their type, i.e., by wrapping the type in the \code{std::atomic}
template. During the compilation, the compiler translates all accesses to such
variables such that the appropriate atomic intrinsic is used for every
access. Our automatically generated header can then insert the appropriate calls
to our replication API automatically by overriding these intrinsics. In summary,
if we use a C++11 compliant compiler, the patching effort can be limited to
modifying the type of all variables that need to be accessed and updated
atomically. In our future work, we plan to extend our current embryonic
implementation in LLVM to automate this process completely.

The replication-enabled libraries can replace their original counterparts, in
which case they will function correctly and with minimal overhead outside the
MVEE context. Alternatively, they can be installed side-by-side with the
original ones, in which case the MVEE will intervene transparently in the
library loading to load the replication- enabled ones. Our solution hence places
a minimal burden on system administrators and users.

\subsection{Interaction with the Kernel}

Synchronization algorithms often rely on the kernel's \code{futex} API to
interact with other threads and processes.  The multi-purpose synchronization
API is exposed through a system call, and is used throughout GNU's
\code{pthreads} library for two reasons. First, some functions use the
\code{FUTEX\_WAIT} operation to block the calling thread until the value stored
at a specified address changes. In the \code{pthread\_mutex\_lock} function,
this operation is used to block the calling thread if the mutex is currently
contended. Other functions such as \code{pthread\_cond\_wait} use the
\code{FUTEX\_WAIT} operation to block until an event occurs. Other functions use
the \code{FUTEX\_WAKE} or \code{FUTEX\_CMP\_REQUEUE} operation to signal threads
that are waiting for the value at the specified address to change. In the
\code{pthread\_mutex\_unlock} function, wake operations are used to wake up
threads that are blocked in a related \code{pthread\_mutex\_lock} call.
Functions such as \code{pthread\_cond\_signal} or
\code{pthread\_cond\_broadcast} use wake operations to signal and wake up
threads that are blocked inside a related \code{pthread\_cond\_wait} call.

A potential issue arises when a thread performs a wake operation for which only
one other thread should be woken up. If more than one thread is waiting in each
replica and the replication agent does not intervene, the kernel might wake up
non-corresponding threads in the replicae. In a slave replica, the woken thread
will then stall indefinitely (i.e., deadlock) at the first atomic op it
encounters.

While this issue can be handled in the replication agents embedded in all
replicae, we chose to tackle the issue from within GHUMVEE's monitor
instead. The monitor allows all replicae to invoke \code{futex} calls but it
allows only the master replicae to actually complete the call. The monitor
manipulates the system call number and arguments for the slave replicae's
\code{futex} calls to have them perform a harmless non-blocking system call
instead (such as \code{sys\_getpid}). The non-blocking call returns immediately
from the kernel, at which point the monitor stalls the slave until the
master's \code{futex} call has also returned. At that point, the monitor
replicates the result of the system call to all slaves and resumes
them. This guarantees that the corresponding threads get woken up in all
replicae. By implementing the logic in the monitor instead of in the agent, we
keep the agent small and fast for its other replication tasks. The additional
overhead of going through the monitor is relatively small, given that it already
intercepts all system calls invocations and returns anyway.

\section{Evaluation}
\label{evaluation}

Our implementation of GHUMVEE supports the i386 and AMD64 architectures for the
GNU/Linux platform but there are no fundamental restrictions to either the
architectures, or the platform: All design options we lifted for GHUMVEE target
applications running on top of an unmodified OS running on a commercial
off-the-shelf multi-core processor.

GHUMVEE has already been tested successfully on desktop
programs~\cite{volckaert2015cloning,volckaert2013ghumvee}. In this paper, we
focus on evaluating the replication of explicitly parallel benchmarks, on the
effort needed to patch libraries and to embed our synchronization replication
agent, and on how the replication buffers might enable attacks. As underlying
diversity scheme to mitigate code reuse attacks, GHUMVEE implements Disjount Code
Layouts (DCL). DCL ensures that any given virtual address points to a valid
executable region in no more than one replica. This policy was demonstrated to
provide effective protection against code reuse
attacks~\cite{volckaert2015cloning}.

\subsection{Embedding the replication agent}
\label{eval_ready}
To evaluate the run-time overhead of GHUMVEE, we used the PARSEC 2.1 benchmark
suite. We did not include the \code{facesim} and \code{canneal} benchmarks
because they contain many data races. For \code{canneal}, this is hardly
surprising as it is based on data race recovery. We applied minor
patches\footnote{At \url{http://ghumvee.elis.ugent.be} our patches, raw data and scripts are available. GHUMVEE will
  be open sourced in Q4 2015.}  to four benchmarks to eliminate data races
or to embed our agent.
\code{ferret} raced on the \code{cnt\_enqueue} and \code{input\_end}
  variables. Additionally, the \code{imagick} library on which \code{ferret} depends contained
  unprotected accesses to the \code{free\_segments} variable.
\code{freqmine} raced on the \code{thread\_begin\_status} variable.
\code{raytrace} used ad hoc synchronization in its \code{AtomicCounter}
  class.
\code{vips} had an unprotected read and write in the \code{gclosure.c} file.

We further applied fixes for bugs which had been reported in the literature or
on the PARSEC web site. We configured \code{fluidanimate} and
\code{streamcluster} benchmarks to use the original pthread-based barriers
rather than the semantically equivalent but less efficient parsec-based
barriers.

On our testing system, the benchmark suite relies on four libraries in which we
embedded our replication agent: \code{glibc 2.19}, \code{libpthread 2.19},
\code{libstdc++ 4.8.2} and \code{libgomp 4.8.2}, the default library versions of
Ubuntu 14.04. Since \code{libpthread} and \code{glibc} are built from the same
source tree, we treat them as one entity when reporting the required patching
effort.

For \code{libgomp} and \code{libstc++}, we leverage the use of compiler
intrisics as discussed in Section~\ref{rdready}. Both libraries support
specialized targets and more generic targets: \code{libstdc++} supports the
so-called \code{i486} and \code{generic CPU} targets, while \code{libgomp}
supports the so-called \code{Linux} and \code{POSIX} targets. We adapted the
makefiles, directory structures, and linker scripts to ensure that the code
targeting the \code{generic CPU} and \code{POSIX} are used instead of the code
in support of the more specific \code{i486} and \code{Linux} targets, thus
ensuring that code relying on compiler intrisics is used instead of code
involving inline assembly. Furthermore, we made sure that the automatically
generated header was included in all relevant source files.  

For each of the libraries, all of this preparation required editing/executing
less than 14 lines of script and source code. The automatically generated header
consists of 131 LoC. In addition, in \code{libgomp}, 2 lines of code needed to
be edited to replace two unprotected load/store operations by atomic ones. So
all in all, a very limited effort was required to prepare these libraries:
2*14+2=30 lines of code needed to be edited manually to prepare the two
libraries that total about 110k LoC. Moreover, this manual editing was limited
to 4 source files out of a total of 673 files.

A considerably larger patching effort was needed to embed our agent in
\code{glibc/libpthread}, because they use ad hoc synchronization throughout and
have many explicit memory barriers and unprotected loads and stores.

Whereas more modern \code{glibc/libpthread} ports like the ARM port use compiler
intrinsics to implement their atomic operations, the AMD64 and i386 ports do
not, presumably because intrinsic support in compilers was not up to par yet
when those ports were developed. Instead, the AMD64 and i386 ports rely on
inline assembler. With today's compiler support for intrinsics, the inline
assembler can be replaced by intrinsics without performance penalty. In
addition, our effort for embedding the replication agent in \code{glibc} would
have been much reduced in case the inline assembler had already been replaced by
the intrinsics. As this is not yet the case, we needed to do the replacement
ourselves. For this purpose, we replaced the x86 version of the
\code{lowlevellock.h} header by the ARM version of that same file. We also
deleted the assembly-based x86-specific versions of many \code{pthread}
functions from the source tree, such that the generic versions of those same
functions are used instead.

This did not suffice, however. Contrary to \code{libgomp} and \code{libstc++},
\code{glibc}'s generic code does not use compiler intrinsics directly. Instead,
\code{glibc} implements its own series of sync ops, of which some map directly
to compiler intrinsics and others do not. So we opted to wrap \code{glibc}'s
sync ops manually, rather than with the automatically generated header. Our
manually constructed wrappers span 211 lines of code. We added an additional 175
lines of code to allow \code{ld-linux}, which is also built from \code{glibc}'s
source tree, to still use the original unwrapped macros as needed by GHUMVEE. We
therefore needed 386 lines of code in total to wrap all synchronization
operations in \code{glibc-libpthread}. In addition, we added approx.\ 261 lines
of code to eliminate data races.

Finally, we added 14 lines in one of \code{glibc}'s linker scripts to expose our
replication agent to other libraries, and added our replication agent
itself. The WoC agent, for example, counts no more than 194 lines of
C code, while the secure WoC agent counts 167 lines of assembly code
and 100 lines of C code.

Excluding the copying and deleting of existing code, as well as our own
replication agent, the source code patching effort to prepare
\code{glibc/libpthread} was limited to 211+175+261+14 = 661 LoC in 60 source
code and build files. Compared to the library's total size of several 100K LoC
spread over several thousand source code files, this effort is still fairly
limited. And all of it can of course be reused for replicating all applications.

Moreover, since version 2.20, a gradual effort is ongoing in the glibc developer
community to replace inline assembler sync ops by their more portable, more
generic, more maintainable counterparts in the form of compiler
intrinsics. Together with the automated support we are developing as mentioned
near the end of Section~\ref{rdready}, this will reduce the required patching
effort significantly in the near future. 

\subsection{Run-time overhead and scalability}
We evaluated our technique on a system with two Intel Xeon E5-2650L processors
with 8 physical cores and 20MB cache each. The system has 128GB of main memory
and runs the AMD64 version of the Ubuntu 14.04 OS. For the sake of
reproducibility, we disabled hyper-threading and all power saving and dynamic
frequency and voltage scaling features. The system runs a Linux 3.13.9 kernel
that was compiled with a 1000Hz tick rate to minimize the monitor's latency in
reacting to system calls. We applied a small optional kernel patch (less than 10
LOC) that adds a variant of the \code{sys\_sched\_yield} system call that
bypasses GHUMVEE.  Other than that, no kernel patches were used. With this small
kernel patch, our agents can efficiently yield the CPU whenever they are waiting
for preceding sync ops to finish replaying in the slave replicae. This patch
improves the performance of our TO and WoC agents in the
\code{dedup} benchmark but has no significant effects elsewhere.

All benchmarks were compiled at
optimization level -O2 using \code{GCC 4.8.2}. The native performance of the benchmarks
was measured using the original, unpatched libraries that shipped with
the OS. GHUMVEE performance was measured using their GHUMVEE-enabled versions.

\begin{figure*}[t!]
\centering
\includegraphics[width=0.98\linewidth]{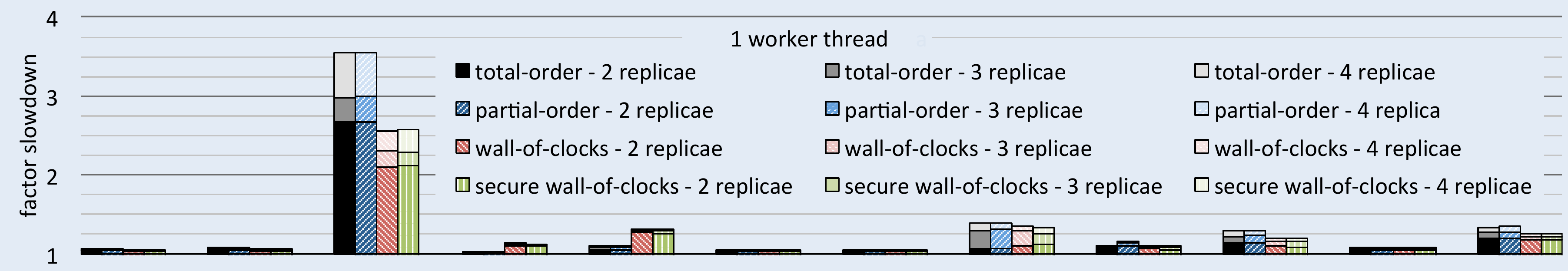}
\vskip 0.15 cm
\includegraphics[width=0.98\linewidth]{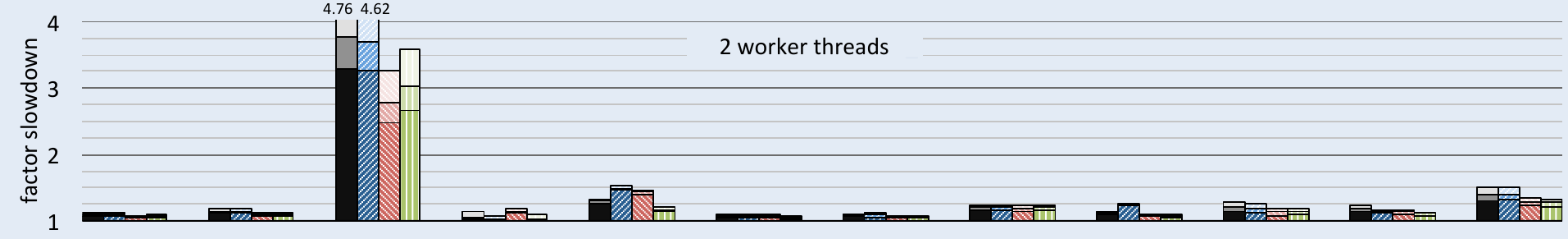}
\vskip 0.15 cm
\includegraphics[width=0.98\linewidth]{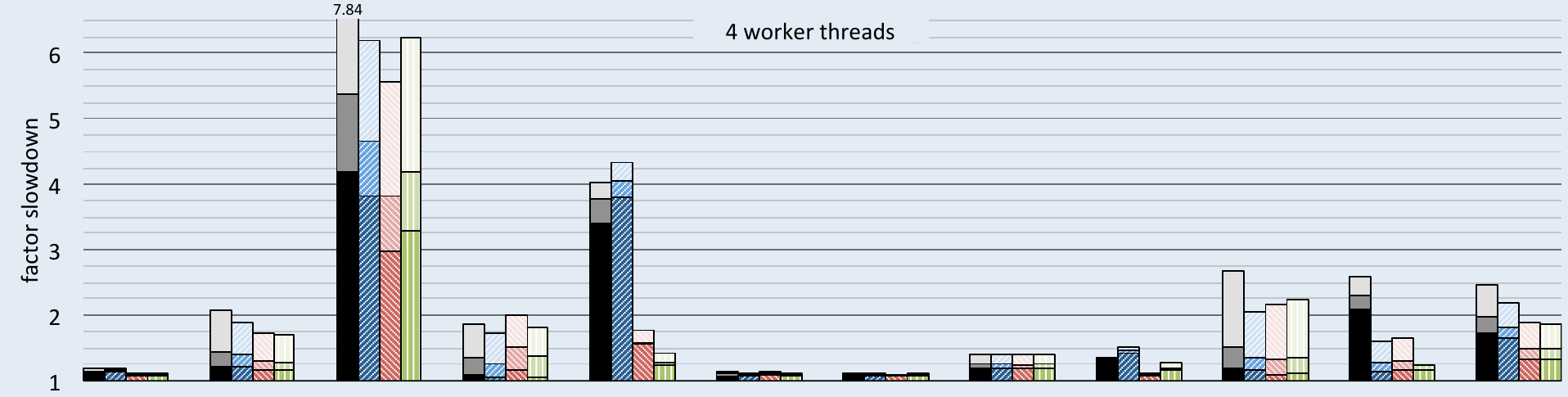}
\vskip 0.15 cm
\includegraphics[width=0.98\linewidth]{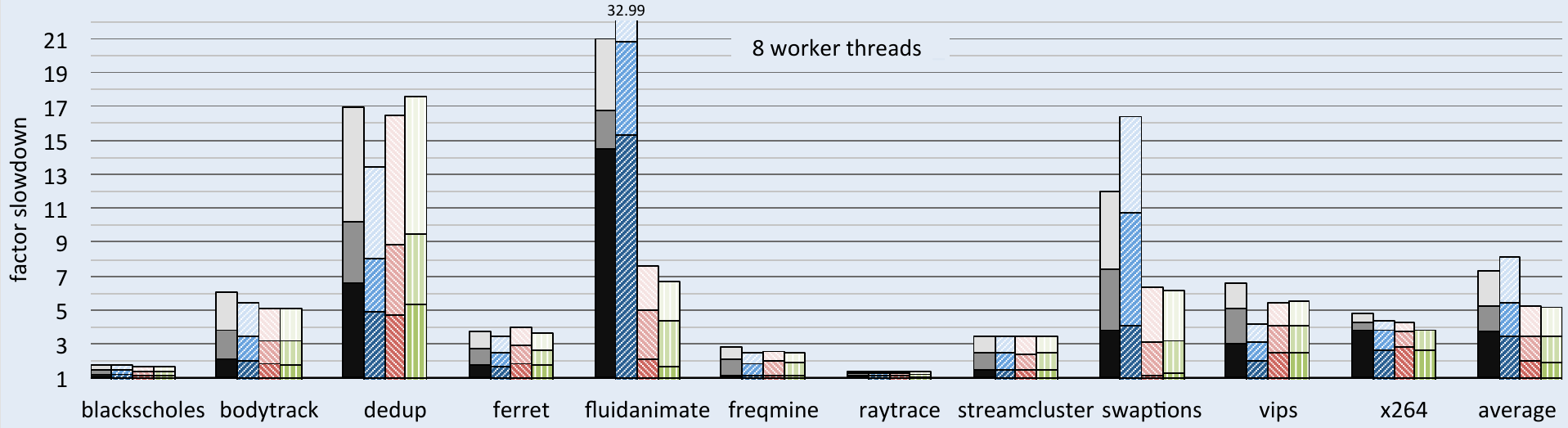}
\caption{GHUMVEE execution time overhead, relative to native execution of Parsec
  2.1 applications for different numbers of worker threads. Each stack for each
  benchmark shows the overhead for 2, 3, and 4 replicae. The four
  stacks per benchmark correspond to the three (non-secured) agents + the secure
  version of the WoC agent.}
\label{overhead}
\end{figure*}

\begin{figure*}[t!]
\centering
\includegraphics[width=14.9cm]{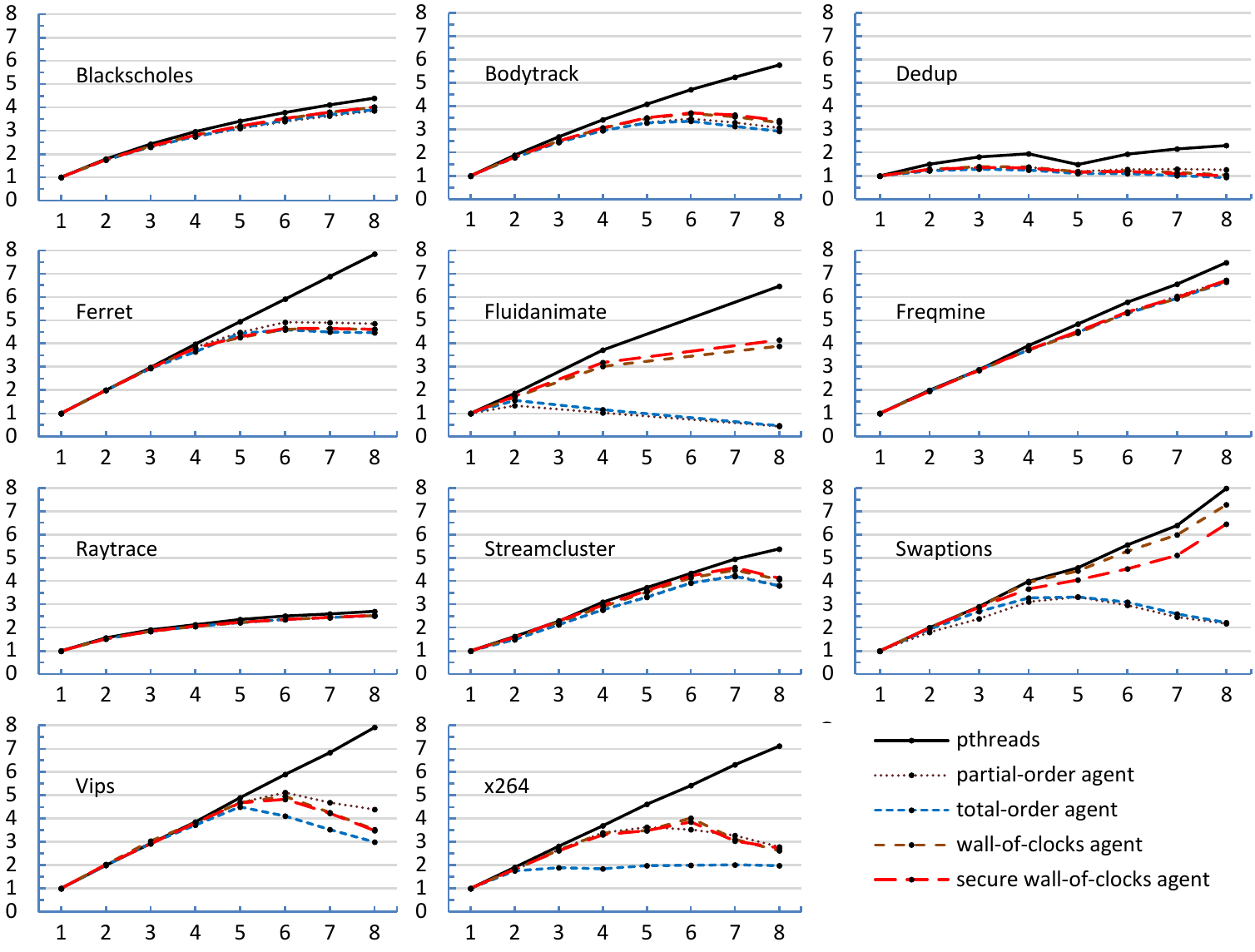}
\caption{Scaling of the native (i.e., pthreads) and replicated benchmarks (for
  two replicae, one master and one slave) with the four different replication
  agents. On the X-axis, the number of worker threads is given. On the Y-axis,
  the performance relative to one worker thread is
  presented. \code{Fluidanimate} only runs when the number of worker threads is
  a power of two.}
\label{scaling}
\end{figure*}

We measured the execution time overhead of our agents by running each PARSEC
benchmark with 1 to 8 worker threads natively as well as in GHUMVEE with 2, 3,
and 4 replicae. Using the native PARSEC input sets, i.e., the largest
standardized set, we ran each measurement five times, of which we omitted the
first to account for I/O-cache warmup. For 1, 2, 4, and 8 worker threads,
Figure~\ref{overhead} presents the benchmarks' execution time as replicated by
GHUMVEE, relative to the native versions. For each agent and for each benchmark,
Figure~\ref{scaling} shows how the native and the replicated execution (for two
replicae) scales with the number of worker threads.

These figures display several trends. Most importantly, with both WoC agents,
many benchmarks (\code{blackscholes}, \code{freqmine}, \code{raytrace},
\code{swaptions}, and even \code{streamcluster)} can be replicated with little
overhead up to 8 worker threads, and in some cases even with 3 or 4
replicae. Other benchmarks (\code{bodytrack}, \code{ferret}, \code{vips},
\code{x264}) can be replicated with little overhead up to 4 worker threads. The
average overhead of the replication remains below 2x for 2 replicae with the WoC
agents, even with 8 worker threads. With more replicae, the overhead clearly
increases. This is of course the result of resource contention of the many
threads over the limited number of cores. For most of the mentioned benchmarks,
it then does not matter too much which agent is used.

For other benchmarks, however, there is a big difference in overhead between the
different agents, and there are several benchmarks for which significantly
larger overheads and bad scaling are observed.

First, regardless of which agent we use, \code{dedup} consistently suffers high
performance penalties. The main contributor to this overhead is the high system
call density in \code{dedup}. When running with 8 worker threads, \code{dedup}
executes over 123k system calls/second.  This density is far greater than in any
other program we have tested so far. In the PARSEC suite itself, the highest
density we have measured besides \code{dedup} was for the \code{vips} benchmark
(20.9k system calls/second for 8 worker threads).  In older benchmark suites
such as SPEC CPU2006, the highest density we have measured was around 1k system
calls/second for \code{403.gcc}. The high overhead in benchmarks with such high
system call densities is unfortunately a fundamental problem of the
\code{ptrace} API on which GHUMVEE and most other security-oriented MVEEs rely
to monitor the behavior of the replicae.

Second, the \code{swaptions} and \code{fluidanimate} benchmarks, which use
fine-grained synchronization, expose a major bottleneck in our PO and
TO agents. Both of these benchmarks use a fork/join threading model and
frequent, fine-grained synchronization. In both of these benchmarks, all worker
threads perform the same tasks and progress at roughly the same pace. While
\code{swaptions} does not use any explicit synchronization in the application
code itself, it does rely heavily on dynamic memory allocation. The dynamic
memory allocator in GNU's \code{libc} uses ad hoc synchronization and lock-free
algorithms to ensure thread safety. Through \code{libc}, \code{swaptions}
executes more than 398M sync ops when running with 5 worker threads. With 8
worker threads, \code{swaptions} performs over 403M sync ops. This corresponds
to 4.2M sync ops per second in the native benchmark with 5 worker threads, and
up to 7.5M sync ops per second in the native benchmark with 8 worker threads.

In \code{fluidanimate}, the situation is even worse. Contrary to
\code{swaptions}, \code{fluidanimate} does invoke our replication agent
directly. With 4 worker threads, \code{fluidanimate} performs over 1.18B sync
ops, which corresponds to over 9.8M sync ops per second in the native
benchmark. These sync ops originate mainly from the \code{pthread\_mutex\_lock}
and \code{pthread\_mutex\_unlock} functions, which are used to acquire or
release one of the 2.31M individual mutexes used in the program. With 8 worker
threads, \code{fluidanimate} performs over 2.35B sync ops, which corresponds to
over 32.9M sync ops per second in the native benchmark. Because the lock and
unlock operations are spread over so many different mutexes, there is very
little contention in the native benchmark.

In GHUMVEE, however, the TO and PO agents create a lot of
contention. Both agents capture the total order of the sync ops in a
single circular buffer. To capture this order, the agents acquire a lock before
executing the original atomic op and do not release this lock until the
operation has been logged in the buffer. These agents therefore effectively
serialize the execution of sync ops in the master replica.

A second problem with these two agents is that all replicae must keep track of
their current position in the buffer. This position must be read before the
processing of each atomic op, and updated after each atomic up. Every time an
update happens on a different core that does not share a cache with the core on
which the previous update was executed, the cache line that contains the current
position in the circular buffer will be invalidated, and hence cause stalls in
the cores' pipelines.

The combination of these two bottlenecks results in poor scaling for \code{swaptions} and \code{fluidanimate}.  In other benchmarks, the
effects of the serialization and the additional cache coherence traffic incurred
by our TO and PO agents are less visible. The main reason is that
the other benchmarks perform much less sync ops than \code{swaptions} and
\code{fluidanimate}. In the other benchmarks, the highest sync op rates occurred
for \code{dedup} and \code{vips}, with 936K and 644K sync ops per second in the
native benchmark resp.

Most importantly, our WoC agents almost completely eliminate the
bottlenecks observed in the \code{swaptions} and \code{fluidanimate}
benchmarks. Only in the \code{vips} benchmark, these agents cannot avoid a
significant serialization overhead when the number of worker threads increases,
because it assigns many unrelated mutexes to the same logical clocks. Thus, for
this specific benchmark, the PO agent outperforms the WoC
agents.

A final trend is that benchmarks that use condition variables do not scale well
beyond 6 worker threads. The \code{bodytrack}, \code{dedup},
\code{ferret}, \code{vips} and \code{x264} benchmarks all rely on condition
variables to signal and to wake up threads. With enough available CPU time, all
of the benchmark's threads can run simultaneously and a thread can be signaled
without resorting to \code{sys\_futex} calls. The five mentioned benchmarks all
have heterogeneous threading models and have more than $n$ threads running
simultaneously (with $n$ the number of worker threads). For that reason, our
machine's 16 cores simply cannot run all threads in 2 or more replicae
simultaneously.

All in all, the WoC agents perform reasonably well. With 4 worker
threads and two replicae, the average slowdown of our MVEE is only 1.33x with
the regular WoC agent and 1.32x with the secured WoC
agent. With the TO and PO agents, the average slowdown is 1.73x
and 1.64x resp.  The high system call overhead in the \code{dedup} benchmark is
the main contributor to the slowdown. In this configuration, the slowdown in
\code{dedup} ranges from 2.98x with the WoC agent to 4.19x with the
TO agent.

With 8 worker threads and two replicae, the average slowdown is much higher. The
slowdown for our TO, PO, WoC and secured WoC
agents is 3.69x, 3.42x, 1.99x, and 1.98x resp. For our TO and PO
agents, the main cause is the introduced serialization and constant cache
invalidations that come with the single circular buffer approach. Our WoC agents eliminate this bottleneck for the most part, but in this
configuration, the lack of resources on our test machine becomes a problem.  The
variations in results for the WoC agents are caused by minor
differences in their implementation. The regular WoC agent accesses
the synchronization replication buffer directly, whereas the secured WoC agent accesses the buffer indirectly, as we explained in
Section~\ref{swoc}. This indirection slightly increases the cache pressure. As
opposed to the regular WoC agent's C implementation however, the
secured WoC agent's hand-written assembler implementation does not
spill any registers to the stack. This optimization slightly reduces the cache
pressure. The combination of these two minor implementation differences slightly
favors the secured WoC agent in terms of performance.

\subsection{Security Evaluation}
All of our replication agents rely on a buffer that is shared between all
replicae.  This buffer is mapped as a read-write memory segment in the master
replica and as a read-only segment in the slave replicae. Intuitively, it might
seem like a security risk to create such a communication channel between the
replicae because it can be used to forward information from the master to the
slave replicae without triggering a monitored RVP. In practice, however, the
security risk is minimal.

In principle, it is possible to launch attacks that cause the master replica to
write arbitrary data into the buffer, but the master replica cannot instruct the
slave replicae to use the arbitrary data in any meaningful way other than to
replay synchronization operations, because the data written into the
synchronization buffer is only read by the replication agent. We have manually audited
our replication agents and verified that they never pass any information they
retrieve from the synchronization buffer on to any other part of the
program. GHUMVEE further ensures that explicit input, i.e., input retrieved from
system calls, is never written into the synchronization buffer.

We do, however, anticipate future MVEE designs in which the MVEE does not
arbitrate all system calls that may retrieve input. For example, the recently
proposed reliability-oriented VARAN handles system call monitoring and input
replication entirely in user space and inside the context and address space of
the replicae~\cite{hosek2015varan}. In such a system, the synchronization buffer
could in theory be used as an uncontrolled communication channel, which might
aid a compromised master replica in mounting an attack on the slave replicae.
Specifically, the compromised master replica could manipulate the return values
of its system calls, thereby instructing the slave replicae to read further
input from the synchronization buffer.

To protect the synchronization replication buffer in this scenario, GHUMVEE
forces the buffer to be mapped at different, randomized addresses in each
replica. A compromised master replica therefore would not know the exact
location of the buffer in the slave replicae and it would have to derive the
location through \textbf{information leakage} or by
\textbf{guessing}. Alternatively, the master replica could try to construct a
\textbf{code reuse attack} that invokes the replication agent's code to read
from the synchronization buffer.

GHUMVEE prevents the latter attack with its DCL. The master replica can therefore not
mount a code reuse attack: He cannot assume that slave replicae have the same
memory layout as the master replica, and if he feeds an address to the replicae
that points to an executable gadget in the master's address space, the slave
will raise an exception when it tries to execute code at the same address.

Guessing the location of the buffer is hard. GHUMVEE currently use
synchronization buffers of $256MiB$, which corresponds to $65536$
memory pages. The AMD64 ABI allows user-space applications to use 48 bits for
memory addressing but excludes the first memory page, i.e., the page that starts
at address $0x0$. Therefore, a user-space application may map up to $2^{36} - 2$
memory pages. The chance to blindly guess the location of a $256MiB$ buffer in
one slave replica is therefore $65536 / (2^{36} - 2)$ or $9,53\cdot10^{-7}$.

Forcing the slave replicae to leak the location of their synchronization buffer
is not trivial either. While the TO, PO and regular
WoC agents do internally store a pointer to the synchronization
buffer, GHUMVEE prevents leakage of the pointer from the slaves to the master
through the buffer by mapping the buffer as read-only in the slaves. As GHUMVEE
intercepts all system calls, it is trivial to prevent a replica from reverting
that memory protection.

Leaking the pointer through other channels is still possible, however. We have
therefore constructed our secured WoC agent, which significantly
reduces the odds of a successful leakage attack. This secured agent does not
store a pointer to the synchronization buffer, but instead accesses the
synchronization buffer only through an indirection via the \texttt{gs} segment,
as explained in Section~\ref{swoc}.

A remaining option to consider is the use of a covert channel between the
replicae, and to use the MVEE as the medium through which the covert channel
communicates.  For example, the replicae can deliberately delay each other by
exploiting the MVEE's lock-step execution mechanism. This mechanism dictates
that certain operations may only be completed when all replicae attempt to
invoke them. The length of the delay can represent information such as
individual bits of a pointer value. While it is easy to write programs that
intentionally set up and exploit such channels, it is not possible to deploy
this technique if the MVEE's protection policy is properly implemented. In
GHUMVEE, DCL prohibits the launch of a code reuse attack (to setup and exploit
the covert channels) in the first place.

In conclusion, we believe that our synchronization replication buffers and
agents are sufficiently hardened against attacks, even in scenarios where the
master replica can forward explicit input to the slave replicae via those
buffers.

\section{Related Work}
\label{Related}

\subsection{MVEEs}
Throughout the last decade, several MVEEs have been presented. Cox et al.
first proposed N-Variant Systems, a kernel-space MVEE~\cite{cox2006n}. Shortly
afterwards, Cavallaro presented a proof-of-concept user-space
MVEE~\cite{cavallaro2007comprehensive}. Salamat et al. then proposed Orchestra,
a more advanced user-space MVEE~\cite{salamat2009orchestra}. More recently,
Hosek and Cadar presented Mx~\cite{hosek2013safe} and
VARAN~\cite{hosek2015varan}, while Maurer and Brumley introduced
Tachyon~\cite{maurer2012tachyon}. The latter three systems are not
security-oriented MVEEs like the former ones, as they aim at safe testing of
experimental software updates, rather than at protecting programs against
exploits. The only multi-threaded applications on which VARAN was tested were
server applications in which none of the system call behavior depends on the
thread synchronization order: Those server benchmark threads perform almost
completely independent computations. By contrast, the system call behavior in
the PARSEC benchmarks, even our data race free versions, depends very much on
the synchronization order. Without replicating and ordering synchronization
events, none of the PARSEC benchmarks can be handled correctly. VARAN's approach
of ordering system calls and signals but not synchronization events, is simply
not a generic, reliable solution. While not reported in detail in this paper, we
also successfully tested GHUMVEE on all benchmarks on which VARAN and all other
mentioned MVEEs were reportedly evaluated. While some of those MVEEs have been
evaluated on multi-process applications (such as older version of Apache),
GHUMVEE is the first to provide active support for non-deterministic,
multi-threaded applications.

\subsection{Deterministic Multithreading}

Deterministic MultiThreading (DMT) systems impose a deterministic schedule on
the execution order of instructions that participate in inter-thread
communication, or a deterministic schedule on the order in which the effects of
those instructions become visible to other threads. Some DMT systems guarantee
determinism only in the absence of data races (\emph{weak determinism}), while
others work even for programs with data races (\emph{strong determinism}).

Some DMT implementations, especially the older ones, rely on custom
hardware~\cite{devietti2009dmp,devietti2011rcdc,hower2011calvin,basu2011karma} or a custom operating
system~\cite{aviram2012efficient,bergan2010deterministic}. Of interest to us, however, are the user-space
software-based approaches~\cite{basile2002preemptive,reiser2006consistent,berger2009grace,liu2011dthreads,merrifield2013conversion,cui2013parrot,olszewski2009kendo,lu2014efficient,devietti2009dmp,bergan2010coredet,zhou2012exploiting}.

Software-based DMT systems come in many flavors but essentially, they all
establish a deterministic schedule by passing a token. We refer to the
literature for an excellent overview of the possible ways to implement the
deterministic schedule, as well as their implications~\cite{segulja2014cost}. In
the remainder of this discussion, we focus on the fundamental reason why DMT
systems are incompatible with MVEEs that run diversified replicae: the timing of
and prerequisites for the deterministic token passing.

Most DMT systems require that all threads synchronize at a global barrier before
they can pass their token. Some of the systems that employ such a global
barrier, insert calls to the barrier function only when a thread executes a
synchronization
operation~\cite{basile2002preemptive,reiser2006consistent,berger2009grace,liu2011dthreads}. This
approach is incompatible with parallel programs in which threads deliberately
wait in an infinite loop for an asynchronous event such as the delivery of a
signal to trigger. Such threads never reach the global barrier.  Other DMT
systems tackle this issue by inserting barriers at deterministic points in the
thread's execution. These deterministic points are based on the number of
executed store instructions~\cite{olszewski2009kendo}, the number of issued
instructions~\cite{zhou2012exploiting} or the number of executed
instructions~\cite{bergan2010coredet,devietti2009dmp}. All of these numbers are
extremely sensitive to small program variations, which makes such systems an ill
fit for use in diversified replicae.

Conversion~\cite{merrifield2013conversion} does not use a global barrier but,
like other DMT systems, it relies on a deterministic token that can only be
passed when threads invoke synchronization operations, which again is
incompatible with parallel programs in which some threads never invoke
synchronization operations. RFDet~\cite{lu2014efficient} uses an optimized
version of the Kendo algorithm~\cite{olszewski2009kendo} to establish a
deterministic synchronization order. Like Kendo however, the order is still
based on the amount of executed instructions in each thread, which makes RFDet
equally sensitive to program variations.

\subsection{Record/Replay}
Record/Replay (R+R) systems capture the order of synchronization operations in
one execution of a program and then enforce the same order in a different
execution. This can happen offline, by capturing the order in a file to be
replayed during a later execution of the same program, or online, by
broadcasting the order directly to another running instance of the program.  In
the absence of data races, R+R systems show many similarities with DMT
techniques that impose weak determinism.

RecPlay is a prime example of an offline R+R system~\cite{ronsse1999recplay}. During
recording, RecPlay logs Lamport timestamps for all \code{pthread}-based
synchronization operations~\cite{lamport1978time}. During subsequent replay sessions,
synchronization operations are forced to wait until all operations with a
earlier timestamp have completed. Because it only enforces the order of
synchronization operations, RecPlay's replication mechanism incurs less overhead
than preexisting techniques that replicate the thread scheduling order or the
order in which interrupts are processed~\cite{audenaert1994interrupt}. Moreover,
RecPlay assigns the same timestamp to non-conflicting synchronization
operations, such that they can also be replayed in parallel.

Loose Synchronization Algorithm (LSA) was one of the first techniques that
adopted R+R for use in fault-tolerant systems~\cite{basile2006active}. LSA designates
one of the nodes as the master node. The master node records the order of all
\code{pthread}-based mutex acquisitions and periodically replicates this order
to the slave nodes. These slave nodes then enforce the same acquisition order on
a per-mutex basis.

More recently, Lee et al. proposed Respec online replay on multi-processor
systems~\cite{lee2010respec}. Oriented towards fault-tolerant execution of
identical replicae, Respec purposely records an unprecise order of
synchronization operations in the master process and speculatively replays that
order in the slave processes. At the end of a replay interval, Respec checks
whether the slaves are still synchronized with the master process by comparing
their state, incl.\ their register contents. If not, it rolls them back. While
recording, Respec maps synchronization variables onto a statically allocated
clock, similarly to our WoC agents. It is doubtful, however, whether
Respec's approach could work in a security-oriented MVEE like ours, in which
diversity in the replicae makes it hard (if not impossible) to detect whether
the replicae have diverged at the end of a replay interval.

Other online R+R techniques rely on custom hardware
support~\cite{basu2011karma}, and hence are not useful for a secure MVEE for
off-the-shelf systems.

\section{Conclusions}
\label{conclusions}

This paper presented how GHUMVEE was extended to become the first
security-oriented MVEE that can replicate parallel programs correctly. We
proposed three replication strategies and implemented four replication agents to
implement them, one of which does so over a secured communication channel.  Our
replication agents are conceptually similar to existing tools, but unlike
existing tools, they fit within the constraints that a security-oriented MVEE
imposes for lock-step monitoring of diversified replicae.

Additionally, we proposed a new strategy to embed a replication agent into
parallel programs, incl. programs that use ad hoc synchronization primitives,
and we evaluated the effort to do so. In the future, we plan to automate this
strategy to a large degree.

We extensively evaluated the effect of our MVEE and our replication agents on
the PARSEC 2.1 benchmarks on the GNU/Linux platform. With our secure
wall-of-clocks agent, the best of the four agents, we achieve an average
slowdown of just 1.32x when running the benchmarks with 4 worker threads and 2
replicae.

\bibliographystyle{IEEEtran}
\bibliography{paper}

\begin{thebibliography}{10}
\providecommand{\url}[1]{#1}
\csname url@rmstyle\endcsname
\providecommand{\newblock}{\relax}
\providecommand{\bibinfo}[2]{#2}
\providecommand\BIBentrySTDinterwordspacing{\spaceskip=0pt\relax}
\providecommand\BIBentryALTinterwordstretchfactor{4}
\providecommand\BIBentryALTinterwordspacing{\spaceskip=\fontdimen2\font plus
\BIBentryALTinterwordstretchfactor\fontdimen3\font minus
  \fontdimen4\font\relax}
\providecommand\BIBforeignlanguage[2]{{%
\expandafter\ifx\csname l@#1\endcsname\relax
\typeout{** WARNING: IEEEtran.bst: No hyphenation pattern has been}%
\typeout{** loaded for the language `#1'. Using the pattern for}%
\typeout{** the default language instead.}%
\else
\language=\csname l@#1\endcsname
\fi
#2}}

\bibitem{PaXASLR}
{PaX Team}, ``Address space layout randomization,''
  \url{http://pax.grsecurity.net/docs/aslr.txt}, 2004.

\bibitem{cowan1998stackguard}
C.~Cowan, C.~Pu, D.~Maier, H.~Hinton, J.~Walpole, P.~Bakke, S.~Beattie,
  A.~Grier, P.~Wagle, Q.~Zhang, \emph{et~al.}, ``Stackguard: Automatic adaptive
  detection and prevention of buffer-overflow attacks,'' in \emph{Proc. 7th
  USENIX Security Symp.}, vol.~81, 1998, pp. 346--355.

\bibitem{PaXWX}
{PaX Team}, ``{PaX} non-executable pages design \& implementation,''
  \url{http://pax.grsecurity.net/docs/noexec.txt}, 2004.

\bibitem{shacham2004effectiveness}
H.~Shacham, M.~Page, B.~Pfaff, E.-J. Goh, N.~Modadugu, and D.~Boneh, ``On the
  effectiveness of address-space randomization,'' in \emph{Proc. ACM Conf.
  Computer and Communications Security}, 2004, pp. 298--307.

\bibitem{richarte2002four}
G.~Richarte \emph{et~al.}, ``Four different tricks to bypass stackshield and
  stackguard protection,'' \emph{World Wide Web}, vol.~1, 2002.

\bibitem{shacham2007geometry}
H.~Shacham, ``The geometry of innocent flesh on the bone: Return-into-libc
  without function calls (on the x86),'' in \emph{Proc. ACM Conf. Computer and
  Communications Security}, 2007, pp. 552--561.

\bibitem{goktas2014out}
E.~Goktas, E.~Athanasopoulos, H.~Bos, and G.~Portokalidis, ``Out of control:
  Overcoming control-flow integrity,'' in \emph{2014 IEEE Symposium on Security
  and Privacy (SP)}, 2014, pp. 575--589.

\bibitem{davi2014stitching}
L.~Davi, D.~Lehmann, A.-R. Sadeghi, and F.~Monrose, ``Stitching the gadgets: On
  the ineffectiveness of coarse-grained control-flow integrity protection,'' in
  \emph{USENIX Security Symposium}, 2014.

\bibitem{carlini2014rop}
N.~Carlini and D.~Wagner, ``Rop is still dangerous: Breaking modern defenses,''
  in \emph{USENIX Security Symposium}, 2014.

\bibitem{goktas2014size}
E.~G{\"o}kta{\c{s}}, E.~Athanasopoulos, M.~Polychronakis, H.~Bos, and
  G.~Portokalidis, ``Size does matter: Why using gadget-chain length to prevent
  code-reuse attacks is hard,'' in \emph{Proc. 23rd USENIX Security Symp.},
  2014, pp. 417--432.

\bibitem{schuster2014evaluating}
F.~Schuster, T.~Tendyck, J.~Pewny, A.~Maa{\ss}, M.~Steegmanns, M.~Contag, and
  T.~Holz, ``Evaluating the effectiveness of current anti-rop defenses,'' in
  \emph{Research in Attacks, Intrusions and Defenses}.\hskip 1em plus 0.5em
  minus 0.4em\relax Springer, 2014, pp. 88--108.

\bibitem{wahbe1994efficient}
R.~Wahbe, S.~Lucco, T.~E. Anderson, and S.~L. Graham, ``Efficient
  software-based fault isolation,'' in \emph{ACM SIGOPS Operating Systems
  Review}, vol.~27, no.~5, 1994, pp. 203--216.

\bibitem{abadi2005control}
M.~Abadi, M.~Budiu, U.~Erlingsson, and J.~Ligatti, ``Control-flow integrity,''
  in \emph{Proceedings of the 12th ACM conference on Computer and
  communications security}, 2005, pp. 340--353.

\bibitem{kuznetsov2014code}
V.~Kuznetsov, L.~Szekeres, M.~Payer, G.~Candea, R.~Sekar, and D.~Song,
  ``Code-pointer integrity,'' in \emph{USENIX Symposium on Operating Systems
  Design and Implementation (OSDI)}, 2014.

\bibitem{larsen2014sok}
P.~Larsen, A.~Homescu, S.~Brunthaler, and M.~Franz, ``Sok: Automated software
  diversity,'' in \emph{2014 IEEE Symposium on Security and Privacy (SP)},
  2014, pp. 276--291.

\bibitem{cox2006n}
B.~Cox, D.~Evans, A.~Filipi, J.~Rowanhill, W.~Hu, J.~Davidson, J.~Knight,
  A.~Nguyen-Tuong, and J.~Hiser, ``N-variant systems: A secretless framework
  for security through diversity,'' in \emph{Proc. 15th USENIX Security Symp.},
  2006, pp. 105--120.

\bibitem{salamat2009orchestra}
B.~Salamat, T.~Jackson, A.~Gal, and M.~Franz, ``Orchestra: intrusion detection
  using parallel execution and monitoring of program variants in user-space,''
  in \emph{Proc. EuroSys Conf.}, 2009, pp. 33--46.

\bibitem{cavallaro2007comprehensive}
L.~Cavallaro, ``Comprehensive memory error protection via diversity and
  taint-tracking,'' Ph.D. dissertation, Univ. Degli Studi Di Milano, 2007.

\bibitem{maurer2012tachyon}
M.~Maurer and D.~Brumley, ``Tachyon: Tandem execution for efficient live patch
  testing.'' in \emph{USENIX Security Symposium}, 2012, pp. 617--630.

\bibitem{hosek2013safe}
P.~Hosek and C.~Cadar, ``Safe software updates via multi-version execution,''
  in \emph{Proceedings of the 2013 International Conference on Software
  Engineering}, 2013, pp. 612--621.

\bibitem{volckaert2013ghumvee}
S.~Volckaert, B.~De~Sutter, T.~De~Baets, and K.~De~Bosschere, ``{GHUMVEE}:
  efficient, effective, and flexible replication,'' in \emph{Proc. Int'l Symp.
  on Foundations and practice of security}, 2013, pp. 261--277.

\bibitem{salamat2008reverse}
B.~Salamat, A.~Gal, and M.~Franz, ``Reverse stack execution in a multi-variant
  execution environment,'' in \emph{Workshop on Compiler and Architectural
  Techniques for Application Reliability and Security}, 2008, pp. 1--7.

\bibitem{chew2002mitigating}
M.~Chew and D.~Song, ``Mitigating buffer overflows by operating system
  randomization,'' 2002.

\bibitem{volckaert2015cloning}
S.~Volckaert, B.~Coppens, and B.~De~Sutter, ``Cloning your gadgets: Complete
  {ROP} attack immunity with multi-variant execution,'' \emph{IEEE Trans. on
  Dependable and Secure Computing}, 2015, to appear.
  DOI:10.1109/TDSC.2015.2411254.

\bibitem{basile2002preemptive}
C.~Basile, Z.~Kalbarczyk, and R.~Iyer, ``A preemptive deterministic scheduling
  algorithm for multithreaded replicas,'' in \emph{Proc. IEEE Int'l Conf.
  Dependable Systems and Networks}, 2002, pp. 149--158.

\bibitem{reiser2006consistent}
H.~Reiser, J.~Domaschka, F.~J. Hauck, R.~Kapitza, and
  W.~Schr{\"o}der-Preikschat, ``Consistent replication of multithreaded
  distributed objects,'' in \emph{Proc. IEEE Symp. Reliable Distributed
  Systems}, 2006, pp. 257--266.

\bibitem{berger2009grace}
E.~Berger, T.~Yang, T.~Liu, and G.~Novark, ``{Grace: safe multithreaded
  programming for C/C++},'' \emph{ACM Sigplan Notices}, vol.~44, no.~10, pp.
  81--96, 2009.

\bibitem{liu2011dthreads}
T.~Liu, C.~Curtsinger, and E.~Berger, ``{DTHREADS: efficient deterministic
  multithreading},'' in \emph{Proc. ACM Symp. on Operating System Principles},
  2011, pp. 327--336.

\bibitem{merrifield2013conversion}
T.~Merrifield and J.~Eriksson, ``Conversion: Multi-version concurrency control
  for main memory segments,'' in \emph{Proc. ACM European Conference on
  Computer Systems (EuroSys)}, 2013, pp. 127--139.

\bibitem{cui2013parrot}
H.~Cui, J.~Simsa, Y.-H. Lin, H.~Li, B.~Blum, X.~Xu, J.~Yang, G.~A. Gibson, and
  R.~E. Bryant, ``Parrot: A practical runtime for deterministic, stable, and
  reliable threads,'' in \emph{Proceedings of the 24th ACM Symposium on
  Operating Systems Principles (SOSP'13)}, 2013, pp. 388--405.

\bibitem{olszewski2009kendo}
M.~Olszewski, J.~Ansel, and S.~Amarasinghe, ``{Kendo: efficient deterministic
  multithreading in software},'' \emph{ACM Sigplan Notices}, vol.~44, no.~3,
  pp. 97--108, 2009.

\bibitem{lu2014efficient}
K.~Lu, X.~Zhou, T.~Bergan, and X.~Wang, ``Efficient deterministic
  multithreading without global barriers,'' in \emph{Proceedings of the 19th
  ACM SIGPLAN Symposium on Principles and Practice of Parallel Programming
  (PPoPP'14)}, 2014, pp. 287--300.

\bibitem{devietti2009dmp}
J.~Devietti, B.~Lucia, L.~Ceze, and M.~Oskin, ``Dmp: deterministic shared
  memory multiprocessing,'' \emph{ACM SIGARCH Computer Architecture News},
  vol.~37, no.~1, pp. 85--96, 2009.

\bibitem{bergan2010coredet}
T.~Bergan, O.~Anderson, J.~Devietti, L.~Ceze, and D.~Grossman, ``{CoreDet}: a
  compiler and runtime system for deterministic multithreaded execution,''
  \emph{ACM SIGARCH Computer Architecture News}, vol.~38, no.~1, pp. 53--64,
  2010.

\bibitem{zhou2012exploiting}
X.~Zhou, K.~Lu, X.~Wang, and X.~Li, ``Exploiting parallelism in deterministic
  shared memory multiprocessing,'' \emph{Journal of Parallel and Distributed
  Computing}, vol.~72, no.~5, pp. 716--727, 2012.

\bibitem{basile2006active}
C.~Basile, Z.~Kalbarczyk, and R.~Iyer, ``Active replication of multithreaded
  applications,'' \emph{IEEE Trans. on Parallel and Distributed Systems},
  vol.~17, no.~5, pp. 448--465, 2006.

\bibitem{lee2010respec}
D.~Lee, B.~Wester, K.~Veeraraghavan, S.~Narayanasamy, P.~M. Chen, and J.~Flinn,
  ``Respec: efficient online multiprocessor replayvia speculation and external
  determinism,'' \emph{ACM SIGARCH Computer Architecture News}, vol.~38, no.~1,
  pp. 77--90, 2010.

\bibitem{basu2011karma}
A.~Basu, J.~Bobba, and M.~D. Hill, ``Karma: scalable deterministic
  record-replay,'' in \emph{Proc. Int'l Conf. on Supercomputing}, 2011, pp.
  359--368.

\bibitem{ronsse1999recplay}
M.~Ronsse and K.~{De Bosschere}, ``{RecPlay: a fully integrated practical
  record/replay system},'' \emph{ACM Trans. on Computer Systems}, vol.~17,
  no.~2, pp. 133--152, 1999.

\bibitem{lamport1978time}
L.~Lamport, ``Time, clocks, and the ordering of events in a distributed
  system,'' \emph{Communications of the ACM}, vol.~21, no.~7, pp. 558--565,
  1978.

\bibitem{torresrojas1999plausible}
F.~J. Torres-Rojas and M.~Ahamad, ``Plausible clocks: constant size logical
  clocks for distributed systems,'' \emph{Distributed Computing}, vol.~12,
  no.~4, pp. 179--195, 1999.

\bibitem{levrouw1994new}
L.~Levrouw, K.~Audenaert, and J.~Van~Campenhout, ``A new trace and replay
  system for shared memory programs based on {Lamport} clocks,'' in \emph{Proc.
  Euromico Workshop on Parallel and Distributed Processing}, 1994, pp.
  471--478.

\bibitem{sysv}
{Linux Programmer's Manual}, ``{svipc(7) - Linux Manual Page},''
  \url{http://man7.org/linux/man-pages/man7/svipc.7.html}.

\bibitem{IntelManual}
Intel, ``Intel 64 and {IA-32} architectures software developer's manual volume
  {2}: Instruction set reference, a-z,'' 2014.

\bibitem{dijkstra1965solution}
E.~Dijkstra, ``Solution of a problem in concurrent programming control,''
  \emph{Communications of the ACM}, vol.~8, no.~9, p. 569, 1965.

\bibitem{lamport1987fast}
L.~Lamport, ``A fast mutual exclusion algorithm,'' \emph{ACM Transactions on
  Computer Systems (TOCS)}, vol.~5, no.~1, pp. 1--11, 1987.

\bibitem{mellor1991algorithms}
J.~M. Mellor-Crummey and M.~L. Scott, ``Algorithms for scalable synchronization
  on shared-memory multiprocessors,'' \emph{ACM Transactions on Computer
  Systems (TOCS)}, vol.~9, no.~1, pp. 21--65, 1991.

\bibitem{Catomic}
P.~Becker \emph{et~al.}, ``Working draft, standard for programming language
  {C++},'' Technical Report, Tech. Rep. N3242=11-0012, 2011.

\bibitem{hosek2015varan}
P.~Hosek and C.~Cadar, ``{VARAN} the unbelievable: An efficient n-version
  execution framework,'' in \emph{Proc. Int'l Conf. on Architectural Support
  for Programming Languages and Operating Systems}, 2015, pp. 339--353.

\bibitem{devietti2011rcdc}
J.~Devietti, J.~Nelson, T.~Bergan, L.~Ceze, and D.~Grossman, ``{RCDC}: a
  relaxed consistency deterministic computer,'' \emph{ACM SIGPLAN Notices},
  vol.~46, no.~3, pp. 67--78, 2011.

\bibitem{hower2011calvin}
D.~R. Hower, P.~Dudnik, M.~D. Hill, and D.~A. Wood, ``Calvin: Deterministic or
  not? free will to choose,'' in \emph{High Performance Computer Architecture
  (HPCA), 2011 IEEE 17th International Symposium on}, 2011, pp. 333--334.

\bibitem{aviram2012efficient}
A.~Aviram, S.-C. Weng, S.~Hu, and B.~Ford, ``Efficient system-enforced
  deterministic parallelism,'' \emph{Communications of the ACM}, vol.~55,
  no.~5, pp. 111--119, 2012.

\bibitem{bergan2010deterministic}
T.~Bergan, N.~Hunt, L.~Ceze, and S.~D. Gribble, ``Deterministic process groups
  in {dOS}.'' in \emph{OSDI}, vol.~10, 2010, pp. 177--192.

\bibitem{segulja2014cost}
C.~Segulja and T.~S. Abdelrahman, ``What is the cost of weak determinism?'' in
  \emph{Proc. Int'l Conf. Parallel architectures and compilation}, 2014, pp.
  99--112.

\bibitem{audenaert1994interrupt}
K.~M. Audenaert and L.~J. Levrouw, ``Interrupt replay: a debugging method for
  parallel programs with interrupts,'' \emph{Microprocessors and Microsystems},
  vol.~18, no.~10, pp. 601--612, 1994.

\end{thebibliography}

\vspace*{-2\baselineskip}
\begin{IEEEbiography}[{\includegraphics[width=1in,height=1.25in,clip,keepaspectratio]{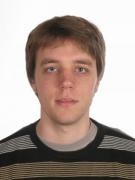}}]{Stijn Volckaert} is Ph.D. student at Ghent University in the Computer Systems Lab. He obtained his BEng. degree in Computer Science from Ghent University's Faculty of Engineering in 2008 and his Msc. degree in Computer Science from Ghent University's Faculty of Engineering in 2010. His research focuses on anti-tampering, multi-variant execution engines, diversity and other mitigations against memory exploits.
\end{IEEEbiography}
\vspace*{-2\baselineskip}
\begin{IEEEbiography}[{\includegraphics[width=1in,height=1.25in,clip,keepaspectratio]{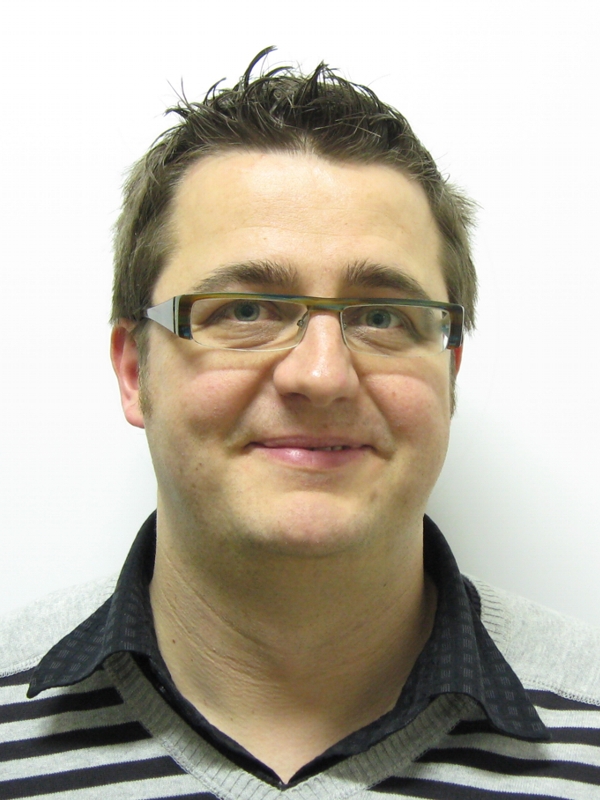}}]{Bjorn De Sutter} is professor at Ghent University in the Computer Systems Lab. He obtained his Msc. and Ph.D. degrees in Computer Science from Ghent University's Faculty of Engineering in 1997 and 2002. His research focuses on the use of compiler techniques to aid programmers with non-functional aspects of their software, such as performance, code size, reliability, and software protection.
\end{IEEEbiography}
\vspace*{-2\baselineskip}
\begin{IEEEbiography}[{\includegraphics[width=1in,height=1.25in,clip,keepaspectratio]{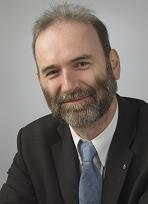}}]{Koen De Bosschere} is professor at the Engineering School of Ghent University, Belgium. His current research interests are binary translation, virtualization, and software protection. He (co-)authored over 170 papers in
journals and conferences. He is the coordinator of HiPEAC, the European Network
of Excellence on High Performance and Embedded Architecture and Compilation and
of the yearly ACACES summer school.
\end{IEEEbiography}
\vspace*{-2\baselineskip}
\begin{IEEEbiography}[{\includegraphics[width=1in,height=1.25in,clip,keepaspectratio]{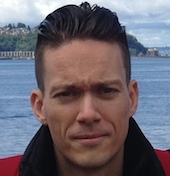}}]{Per Larsen} leads Immunant, Inc. - a security-oriented spin-off from the University of California, Irvine. His research interests include information security, compilation, and optimization. He has a PhD degree from the Technical University of Denmark.
\end{IEEEbiography}

\end{document}